\documentclass[11pt]{article}

\usepackage[a4paper,margin=1in]{geometry}
\usepackage{amsmath,amssymb,bm}
\usepackage{graphicx}
\usepackage{booktabs}
\usepackage{tabularx}
\usepackage{longtable}
\usepackage{array}
\usepackage{caption}
\usepackage{subcaption}
\usepackage{float}
\usepackage{authblk}
\usepackage{enumitem}
\usepackage{hyperref}

\graphicspath{{figures/}}

\setlist[enumerate]{leftmargin=2em}

\newcolumntype{Y}{>{\raggedright\arraybackslash}X}
\newcolumntype{C}{>{\centering\arraybackslash}X}

\title{A Novel Method for Differential-Algebraic Dynamic Model Discovery in Power Systems: An LLM-Based Multi-Agent Collaborative Framework}
\author[1]{Xinming Wang}
\author[1]{Fan Tang}
\author[1]{Yingli Wei}
\author[2]{Yakun He}
\author[1]{Zhe Liu}
\author[3]{Ping Jiang}
\author[3]{Haoyu Wu}
\author[4]{Zihan Guo\textsuperscript{*}}
\author[5]{Chao Shen\textsuperscript{*}}
\affil[1]{Xiong'an New Area Power Supply Company, State Grid Hebei Electric Power Co., Ltd., Baoding 071699, Hebei Province, China}
\affil[2]{State Grid Hebei Electric Power Co., Ltd., Shijiazhuang 050011, Hebei Province, China}
\affil[3]{Mininglamp Technology, Beijing, China}
\affil[4]{School of Advanced Manufacturing and Robotics, Peking University, Beijing 100871, China}
\affil[5]{College of Control Science and Engineering, Zhejiang University, Hangzhou 310027, Zhejiang Province, China}
\date{}

\begin{document}

\maketitle

\noindent\textsuperscript{*}Correspondence: Zihan Guo (2401213194@stu.pku.edu.cn), Chao Shen (12232047@zju.edu.cn)

\begin{abstract}
With large-scale integration of emerging power electronic devices represented by grid-forming inverters, power system dynamics increasingly exhibit strong nonlinearity, multi-timescale coupling, and black-box control logic. These features hinder conventional parameter identification requiring known model structures and structure identification based on predefined function libraries, making complete differential-algebraic dynamic model recovery difficult under weak prior information. To address this challenge, this paper proposes an LLM-based multi-agent collaborative framework for differential-algebraic dynamic model discovery in power systems. It integrates heterogeneous exploratory agents, individual candidate model memories, parameter fitting and evaluation, and a coordinator agent. Under unified measurement-data constraints, agents generate candidate equation structures in parallel, while candidates are optimized, evaluated, retained, and summarized to provide closed-loop search guidance. The task is decomposed into differential equation structure discovery and algebraic closure discovery, enabling joint recovery of state dynamics, algebraic constraints, and key intermediate variables with incomplete prior information. Case studies on synchronous generators and grid-forming inverters show that the proposed method outperforms single-agent LLM-based discovery and conventional symbolic regression in reconstruction accuracy, generalization, search efficiency, and noise robustness. In the generator case, OOD MAPE reaches 0.19\%; in the inverter case, discovery time is reduced by 25.7\% compared with the single-agent LLM baseline.
\end{abstract}


\section{Introduction}

\subsection{Background and Motivation}

With the increasing penetration of renewable energy, inverter-based resources, power electronic interface devices, electric vehicles, and diversified control strategies, modern power systems are undergoing a fundamental transition from dynamics dominated by electromechanical equipment to dynamics dominated by power electronics. Compared with conventional systems mainly governed by synchronous machines and explicit network equations, emerging power systems exhibit stronger nonlinearity, time varying characteristics, and multi time scale coupling among physical networks, converter dynamics, and embedded control loops \cite{ref1}. These features make conventional modelling approaches that rely on complete mechanistic priors increasingly difficult to apply.

This challenge is particularly evident for inverter-based resources and grid forming inverters. Their dynamic responses are determined not only by circuit parameters, but also by inner control loops, outer power or voltage regulation, virtual inertia emulation, damping control, operating mode switching, and software implemented protection logic. In practical applications, many of these control laws are proprietary, partially documented, black box, or time varying. As a result, it is often difficult to obtain complete differential equations, algebraic constraints, and intermediate control variables directly from manufacturers or first principles. Under such conditions, power system dynamic modelling is no longer limited to parameter calibration, but becomes a model discovery problem under incomplete prior information.

Meanwhile, the deployment of phasor measurement units and other wide area monitoring devices makes it possible to record power system dynamic responses with higher temporal resolution. These measurements provide a data basis for extracting dynamic laws and recovering executable models from practical system responses. The objective is not only to reproduce measured trajectories, but also to recover physically meaningful equation structures, variable closure relations, and key intermediate variables that can support dynamic simulation, stability assessment, and control analysis.

\subsection{Literature Review and Research Gap}

Existing studies on power system dynamic modelling and identification can generally be divided into parameter identification and structure identification. Parameter identification methods usually assume that the model structure is known and estimate unknown parameters from disturbance responses, time domain simulation data, or synchronized measurements. Least squares-based methods have been widely used because of their simple mathematical form and convenient implementation. For example, they have been applied to AC network admittance estimation \cite{ref1}, dynamic model parameter estimation \cite{ref4}, measured response calibration \cite{ref5}, and dynamic load modelling \cite{ref7}. However, due to the nonlinear nature of power system dynamic models, direct least squares-based identification may suffer from local convergence, sensitivity to initial values, and high computational cost, especially when numerical differentiation is involved \cite{ref6}. Kalman filter-based identification methods combine model prediction and measurement correction in a recursive framework, making them suitable for online estimation and noisy measurement environments. A cubature Kalman filter has been used to estimate the parameters of a wind power system containing a synchronous generator and a voltage source converter \cite{ref8}. Extended Kalman filters have also been applied to transmission line parameter identification from measurement data \cite{ref9}. To improve mechanical parameter identification under thrust coefficient fluctuations, an equivalent mechanical parameter model and a hybrid adaptive extended Kalman filter based online identifier were developed in \cite{ref10}. These methods have advantages in recursive updating and noise processing \cite{ref11}. However, they still require a predefined state space model. When the model structure, algebraic variables, or intermediate control variables are unknown, Kalman filtering can update parameters but cannot recover missing equations or closure relations.

Data driven parameter identification methods further improve the flexibility of nonlinear parameter estimation. Artificial neural networks have been introduced for dynamic load parameter identification by learning the mapping between measured responses and model parameters \cite{ref12}. Genetic algorithms have been used to construct identification models for automatic voltage regulator parameter estimation \cite{ref13}. More recently, dynamic spatio-temporal adaptive graph neural networks have been proposed for power grid line parameter identification, showing the potential of graph learning in complex networked systems \cite{ref14}. These methods reduce the dependence on repeated solutions of complex physical models and provide useful tools for nonconvex parameter estimation. However, their performance depends strongly on whether the training data sufficiently covers typical operating conditions and disturbance patterns. More importantly, they still focus mainly on parameter learning within a given model form and cannot directly discover unknown differential equations, algebraic constraints, or hidden intermediate variables.

Different from parameter identification, structure identification aims to recover governing equation forms directly from data. Symbolic regression searches for mathematical expressions in a candidate expression space and has been widely studied for discovering differential equations from observational data \cite{ref15}. In power and energy systems, related methods have been used to identify battery degradation processes and extract physically meaningful degradation models from data \cite{ref16}. Sparse identification of nonlinear dynamics further identifies governing equations by selecting active terms from a predefined library, and it has been used to extract reduced order models of grid dynamics \cite{ref17}. Nevertheless, these methods strongly depend on the completeness and correctness of predefined function libraries. If the true physical terms are not included, the identified model may be structurally biased. If the library is too large, the search space expands rapidly, and the model becomes more sensitive to noise and spurious correlations \cite{ref18}. Moreover, fixed library methods are not well suited for power system models involving differential algebraic coupling, missing algebraic variables, and controller induced closure relations.

Recent advances in large language models provide new opportunities for scientific model discovery by combining scientific priors, physical intuition, mathematical expression generation, and executable code synthesis. Representative methods such as LLM-SR formulate equation discovery as a program search process and iteratively improve candidate expressions through generation, parameter fitting, and evaluation \cite{ref19}. Compared with conventional symbolic regression, this strategy improves flexibility and search efficiency. However, LLM-SR mainly targets explicit formula discovery and does not explicitly handle differential algebraic coupling, implicit intermediate variables, or closure constraints commonly found in power system dynamic models. For power system applications, LLM-DMD further introduces LLMs into dynamic model discovery by sequentially discovering differential and algebraic equations and extending missing variables when structural evaluation stagnates \cite{ref18}. Nevertheless, its discovery process still relies on a single LLM and a single reasoning trajectory, where candidate generation, structural correction, variable extension, and search direction updating are dominated by the same model.

Consequently, the search may suffer from limited candidate diversity, fixed inductive bias, local structural paths, and accumulated errors in variable extension.

\subsection{Contributions}

To address the above challenges, this paper proposes an LLM-based multi-agent collaborative framework for power system dynamic model discovery. The framework contains multiple heterogeneous exploratory agents, agent specific candidate model memories, a parameter fitting and evaluation module, and a coordinator agent. Under unified measurement data constraints, different agents generate candidate equation structures in parallel. These candidates are then optimized, evaluated, retained, and summarized to form closed loop search guidance. In this way, local structural exploration, candidate memory retention, parameter optimization, and cross agent information induction are integrated into a unified discovery process.

The main contributions of this study are summarized as follows:

\noindent\textbf{1) A multi-agent collaborative discovery framework is proposed for power system differential algebraic dynamic model discovery.} Different from single trajectory LLM-based discovery, the proposed framework introduces multiple heterogeneous agents to search candidate model structures in parallel under the same measurement data constraints. This design increases candidate diversity, reduces the risk of premature convergence, and improves global exploration capability for complex dynamic model discovery.

\noindent\textbf{2) A candidate memory and coordinator scheduling mechanism is designed to support stable collaborative search.} High quality candidate structures are retained in agent specific memories after unified numerical evaluation. The coordinator agent further summarizes common structural patterns, potential missing variables, and useful search directions from the preferred candidates of different agents. This enables information sharing among multiple search trajectories while preserving structural diversity.

\noindent\textbf{3) A two stage discovery strategy is developed for differential equation recovery and algebraic closure discovery.} The proposed framework adapts LLM-based expression generation to the differential algebraic characteristics of power system dynamic models. Missing algebraic variables, intermediate control variables, and closure relations can therefore be progressively recovered under weak prior information, making the discovered model more suitable for dynamic simulation and system level analysis.

\noindent\textbf{4) The proposed method is evaluated on both electromechanical and power electronic dynamic objects.} A synchronous generator model in the IEEE 39 bus system and a VSG based grid forming inverter model are used as representative test cases. LLM-DMD \cite{ref18}, LLM-SR \cite{ref19}, and SINDy \cite{ref17} are adopted as benchmark methods to assess ID/OOD reconstruction accuracy, discovery efficiency, noise robustness, and system level substitution simulation.

The remainder of this paper is organized as follows. Section~2 defines the dynamic model discovery problem under incomplete prior information. Section~3 presents the proposed LLM-based multi-agent discovery framework. Section~4 reports simulation studies on synchronous generator and grid forming inverter models, including robustness and system-level substitution results. Section~5 concludes the paper and discusses future work.

\section{Problem Definition}
\subsection{Definition of Power System Dynamic Processes}

The dynamic process of a power system under disturbances can be formulated as a set of differential-algebraic equations (DAEs):
\begin{equation*}
\dot{\mathbf{x}} = f(\mathbf{x}, \mathbf{y}, \mathbf{u}; \boldsymbol{\theta}_f),
\tag{1}
\end{equation*}
\begin{equation*}
\mathbf{0} = g(\mathbf{x}, \mathbf{y}, \mathbf{u}; \boldsymbol{\theta}_g).
\tag{2}
\end{equation*}
Here, $\mathbf{x}\in\mathbb{R}^{n_x}$ is the differential state vector, $\mathbf{y}\in\mathbb{R}^{n_y}$ is the algebraic variable vector, $\mathbf{u}\in\mathbb{R}^{n_u}$ is the exogenous input or disturbance vector, $\dot{\mathbf{x}}$ is the time derivative of $\mathbf{x}$, $\boldsymbol{\theta}_f$ and $\boldsymbol{\theta}_g$ are the unknown parameter vectors in the differential and algebraic equations, respectively, $f$ is the state evolution function, and $g$ is the algebraic constraint function.

\subsection{Definition of the Modelling Task}

This paper considers dynamic model discovery under incomplete prior information. In this setting, partial physical knowledge may be available, but the complete equation structure, algebraic constraints, and intermediate variables are not fully known. The task is therefore to recover an executable dynamic model from measured time series data rather than only calibrating parameters of a predefined model.

The available observational dataset is formulated as:
\begin{equation*}
\mathcal{D}=\{(t_k,\mathbf{x}^{\mathrm{obs}}_k,\mathbf{u}_k,\mathbf{z}_k)\}_{k=1}^{N_s}.
\tag{3}
\end{equation*}
where $\mathcal{D}$ is the observational dataset, $k$ is the sample index, $N_s$ is the total number of samples, $t_k$ is the $k$-th sampling instant, $\mathbf{x}^{\mathrm{obs}}_k$ is the available state observation at $t_k$, $\mathbf{u}_k$ is the corresponding exogenous input or disturbance, and $\mathbf{z}_k$ is other measurable outputs or constraint-related quantities.

Based on $\mathcal{D}$, the objective is to jointly discover the differential equation structure, the algebraic equation structure, the associated parameters, and the variable set required for model closure. This process does not assume a fixed basis-function library or a complete set of intermediate variables in advance. The model discovery objective is expressed as:
\begin{equation*}
\widehat{\mathcal{G}}
=
\arg\min_{\mathcal{G}}
\mathcal{J}\!\left(
\mathcal{G};\mathcal{D}
\right),
\qquad
\mathcal{G}=
\left(
f,g,\boldsymbol{\theta}_f,\boldsymbol{\theta}_g,\mathcal{V}
\right).
\tag{4}
\end{equation*}
Here, $\widehat{\mathcal{G}}$ is the discovered dynamic model, $\mathcal{J}$ is the evaluation objective that measures trajectory reconstruction error, and $\mathcal{V}$ is the variable set required for model closure.

Consistent with the weak prior setting of LLM-DMD, only the differential state vector $\mathbf{x}$ is assumed to be known before discovery. The algebraic variables and potential intermediate variables are not completely specified, but are progressively extended during the search process. Therefore, the differential equation and algebraic equation must be identified jointly to obtain a complete DAE model.

\begin{figure}[H]
\centering
\includegraphics[width=\linewidth]{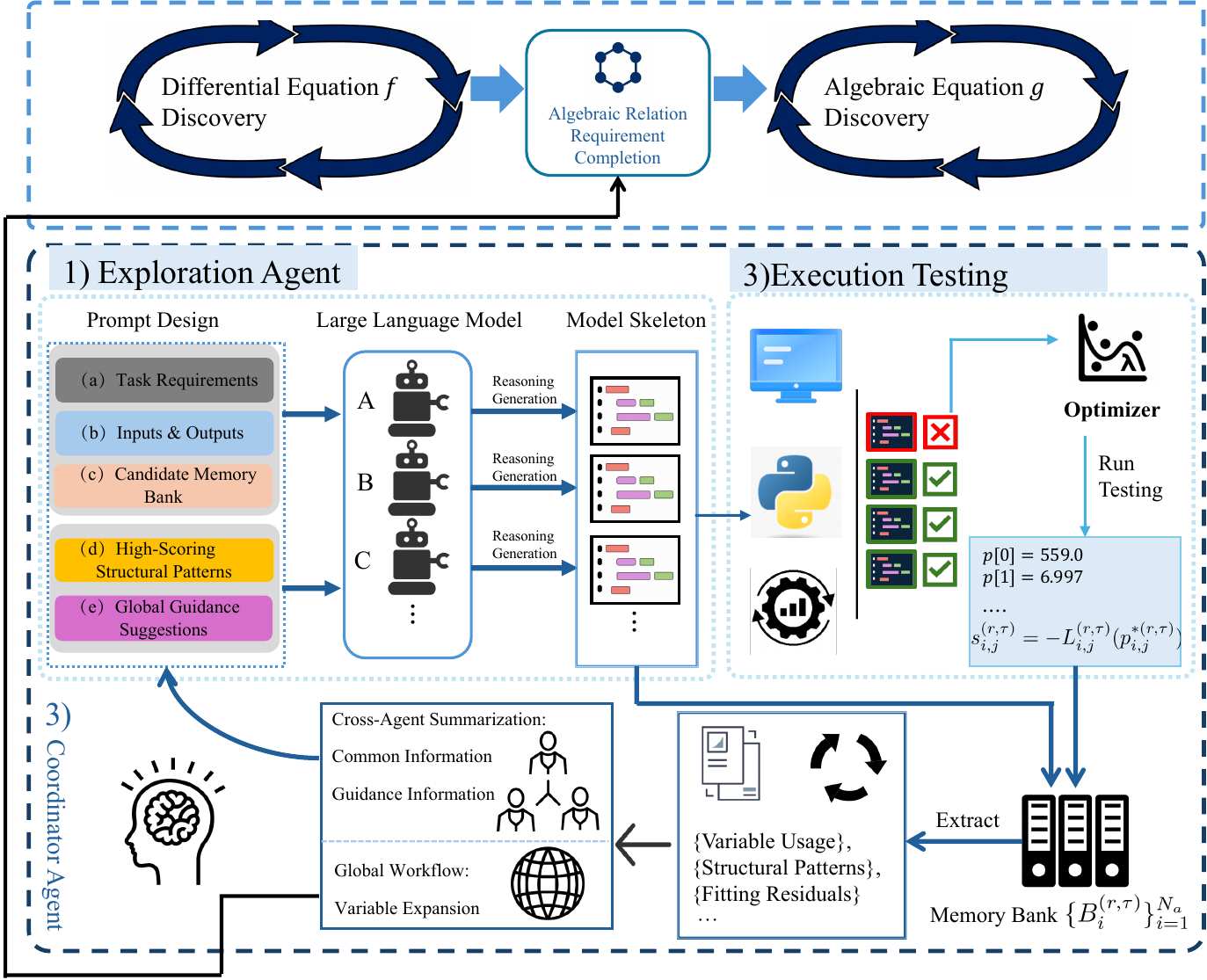}
\caption{Overall flowchart of the multi-agent collaborative adaptive discovery framework.}
\end{figure}

As shown in Fig.~1, the proposed framework follows a two stage discovery process. Differential equation discovery is first performed to identify the state evolution structure, and algebraic closure discovery is then used to complete missing variable relations and constraint equations. Multiple exploratory agents generate candidate model structures in parallel under the same data constraints. The generated candidates are then tested, optimized, and stored in candidate memories. The coordinator agent summarizes cross agent information and broadcasts global search guidance for subsequent iterations. Through this process, candidate generation, parameter determination, memory retention, and variable extension are integrated into the framework.

\section{Method Overview and Framework Components}
\subsection{General Idea}

For the power system dynamic model discovery problem described above, in which the structure is unknown and the variable set may be incomplete, this paper constructs a multi-agent collaborative adaptive discovery framework. The overall process is shown in Fig.~1. For unified notation, the proposed framework is denoted as
\begin{equation*}
\mathcal{F}=(\mathcal{M},\mathcal{B},\mathcal{E},\Omega),
\tag{5}
\end{equation*}
where $\mathcal{M}=\{M_i\}_{i=1}^{N_a}$ denotes the multi-agent structural search unit, which is responsible for generating candidate equation structures under a given search context; $\mathcal{B}=\{B_i\}_{i=1}^{N_a}$ denotes the candidate model memory system, which is responsible for retaining historically preferred candidates of each exploratory agent; $\mathcal{E}$ denotes the parameter determination and candidate evaluation mechanism, which is responsible for converting candidate structural skeletons into comparable models; and $\Omega$ denotes the coordinator agent, which is responsible for cross-agent information induction and global process scheduling. These four components jointly constitute the basic implementation carrier of the proposed method.

In overall operation, the proposed framework iterates around four key links: candidate structure generation, parameter determination, historical retention, and broadcast feedback. For iteration round $r$ and stage $\tau$, the exploratory agent set $\mathcal{M}=\{M_i\}_{i=1}^{N_a}$ generates candidate structure sets $\mathcal{H}_i^{(r,\tau)}$ under the current search context. Then, after processing by the parameter determination and candidate evaluation mechanism $\mathcal{E}$, candidate structures are converted into comparable candidate models and written into the corresponding candidate model memory $B_i^{(r,\tau)}\in\mathcal{B}$ according to scoring results. To prevent the coordinator from intervening too frequently in local search and thereby weakening the structural diversity of parallel exploration, this paper does not update the broadcast summary after every iteration, but adopts a fixed-period broadcast mechanism. Let the broadcast period be $T_b$. When the iteration round satisfies $r\bmod T_b=0$, the coordinator agent $\Omega$ reads the high-quality candidate records in $\{B_i^{(r,\tau)}\}_{i=1}^{N_a}$, forms a global broadcast summary $\mathcal{C}^{(r,\tau)}$, and feeds it back to all exploratory agents. In other rounds, each exploratory agent continues to advance local search only according to its local memory and the most recent broadcast summary. Thus, local structural search, numerical evaluation, historical retention, and cross-agent collaboration form a closed loop within a unified framework.

To keep the method description consistent with the DAE form in the previous problem definition, this paper divides the whole discovery process into a structural discovery process for the differential equation $f(\cdot)$ and a closure discovery process for the algebraic equation $g(\cdot)$, and denotes the current stage indicator as
\begin{equation*}
\tau\in\{f,g\},
\tag{6}
\end{equation*}
where $\tau=f$ denotes the discovery process around the differential equation $\dot{\mathbf{x}}=f(\mathbf{x},\mathbf{y},\mathbf{u};\boldsymbol{\theta}_f)$, and $\tau=g$ denotes the discovery process around the algebraic equation $\mathbf{0}=g(\mathbf{x},\mathbf{y},\mathbf{u};\boldsymbol{\theta}_g)$. Considering that, under initial conditions, the state variables and input variables can be directly obtained while the algebraic variables and intermediate variable set are usually incomplete, the framework first searches for differential structures on the current variable representation. As the search proceeds, if high-quality candidates continuously expose new variable requirements, the process further enters the closure discovery stage for $g(\cdot)$ to supplement the algebraic constraints among related variables. Variable requirements exposed by the former process are supplemented and closed in the latter process, thereby forming a sequentially progressive discovery mechanism.

In iteration round $r$ and stage $\tau$, the variable space currently available for search is denoted as
\begin{equation*}
\mathcal{V}^{(r,\tau)}=\mathcal{X}\cup\mathcal{U}\cup\mathcal{Z}^{(r,\tau)},
\tag{7}
\end{equation*}
where $\mathcal{X}$ denotes the set of differential state variables known a priori, $\mathcal{U}$ denotes the set of exogenous inputs, and $\mathcal{Z}^{(r,\tau)}$ denotes the algebraic variables and intermediate variables that have been included in the current stage. As the discovery process advances, $\mathcal{Z}^{(r,\tau)}$ is progressively expanded under the induction and scheduling of the coordinator agent, so that the search process can continue on a continuously improved variable representation. Finally, the model result output by the proposed framework can be uniformly written as
\begin{equation*}
\widehat{\mathcal{G}}=
\left(
\hat{f},
\hat{g},
\hat{\boldsymbol{\theta}}_f,
\hat{\boldsymbol{\theta}}_g,
\hat{\mathcal{V}}
\right),
\tag{8}
\end{equation*}
where $\hat{f}$ and $\hat{g}$ denote the finally identified differential equation structure and algebraic equation structure, respectively; $\hat{\boldsymbol{\theta}}_f$ and $\hat{\boldsymbol{\theta}}_g$ denote the corresponding parameter estimation results; and $\hat{\mathcal{V}}$ denotes the finally confirmed model variable set.

\subsection{Multi-Agent Structural Search Unit}

The multi-agent structural search unit is the core component responsible for candidate equation generation in the proposed framework. Considering that a single large language model is often limited by a fixed knowledge boundary and reasoning bias in the structural discovery process, and is therefore prone to repeated correction within a limited neighborhood rather than continuously expanding the effective hypothesis space, this paper introduces a parallel search mechanism composed of multiple heterogeneous exploratory agents to improve candidate structure diversity and global search capability. The exploratory agent set is denoted as
\begin{equation*}
\mathcal{M}=\{M_i\}_{i=1}^{N_a},
\tag{9}
\end{equation*}
where $N_a$ is the total number of exploratory agents and $M_i$ denotes the $i$-th exploratory agent. During the entire search process, the agents are mutually independent. They do not exchange candidate structures or share intermediate reasoning results, but advance candidate model generation along their respective search trajectories under unified data constraints.

In iteration round $r$ and stage $\tau$, the search of exploratory agent $M_i$ is built on the currently available variable space $\mathcal{V}^{(r,\tau)}$. To uniformly represent its search input, define
\begin{equation*}
\mathcal{I}_i^{(r,\tau)}
=
\left(
\mathcal{D},
\mathcal{V}^{(r,\tau)},
B_i^{(r,\tau)},
\Pi_i^{(\tau)}
\right),
\tag{10}
\end{equation*}
where $\mathcal{D}$ is the observational dataset used for structure discovery, $\mathcal{V}^{(r,\tau)}$ is the variable set allowed to be called in the current stage, $B_i^{(r,\tau)}$ is the candidate model memory corresponding to the $i$-th agent in this stage, and $\Pi_i^{(\tau)}$ denotes the task constraints and generation settings of the agent in the current stage. These four types of information jointly define the search range of the agent in this iteration. Specifically, the dataset $\mathcal{D}$ provides the numerical basis for structural identification, the variable space $\mathcal{V}^{(r,\tau)}$ determines the physical quantities that candidate equations can call, the candidate model memory $B_i^{(r,\tau)}$ provides local historical reference for structural search, and the task setting $\Pi_i^{(\tau)}$ ensures consistency of candidate generation targets under different stages.

In the concrete implementation, $\Pi_i^{(\tau)}$ is instantiated using a structured prompt template. Its core elements include role and task boundary constraints, declaration of the current available variable space, injection of the coordinator broadcast summary, variable extension output specifications, and target-function interface definitions corresponding to stage $\tau$. By organizing these elements in a unified template, interface consistency can be maintained among different exploratory agents while preserving heterogeneity in generation strategies. An example prompt template is shown in Fig.~2.

\begin{figure}[H]
\centering
\includegraphics[width=0.98\linewidth,height=0.84\textheight,keepaspectratio]{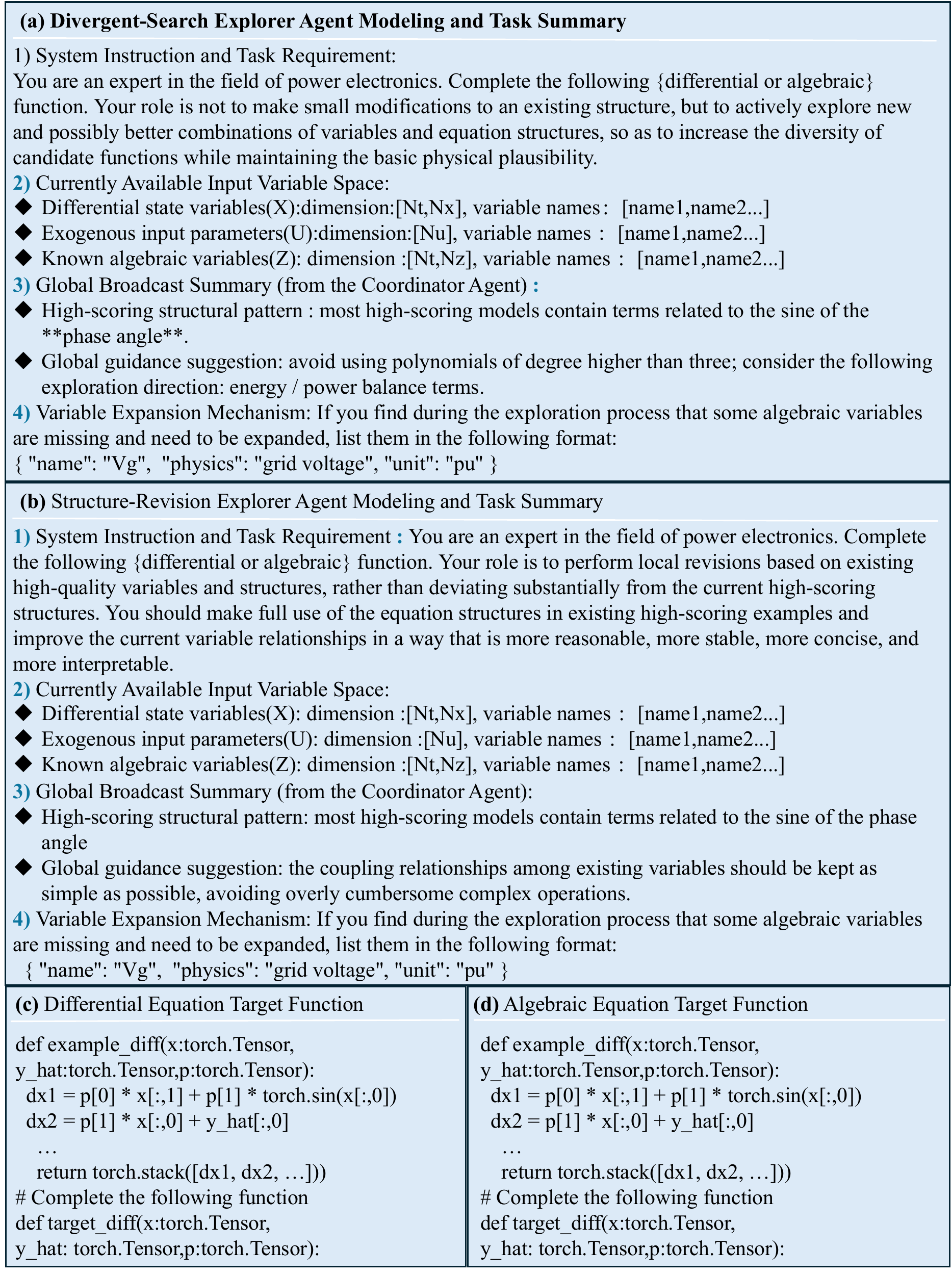}
\caption{Example prompt template for exploratory agents, including divergent exploration and structural correction tasks.}
\end{figure}

Given input $\mathcal{I}_i^{(r,\tau)}$, exploratory agent $M_i$ outputs a set of candidate equation structures, denoted as
\begin{equation*}
\mathcal{H}_i^{(r,\tau)}
=
\left\{
h_{i,j}^{(r,\tau)}
\right\}_{j=1}^{\lambda},
\tag{11}
\end{equation*}
where $\lambda$ denotes the number of candidates generated by the $i$-th agent in iteration round $r$ and stage $\tau$, and $h_{i,j}^{(r,\tau)}$ denotes the $j$-th candidate structure. Thus, the local generation process of the agent can be abstractly represented as
\begin{equation*}
M_i:\mathcal{I}_i^{(r,\tau)}\mapsto \mathcal{H}_i^{(r,\tau)}.
\tag{12}
\end{equation*}
Here, $h_{i,j}^{(r,\tau)}$ does not correspond to a final directly acceptable model, but is a candidate equation expression that remains to be numerically tested. Under different stages, this candidate structure has different meanings: when $\tau=f$, $h_{i,j}^{(r,\tau)}$ corresponds to a candidate differential relation; when $\tau=g$, $h_{i,j}^{(r,\tau)}$ corresponds to a candidate algebraic constraint. For unknown coefficients involved in the candidate structure, this paper retains them uniformly as parameter placeholders during structure generation, rather than letting the exploratory agent directly specify numerical values. The corresponding parameter tuning process is further described in the subsequent parameter determination and candidate model memory maintenance mechanism.

It should be noted that the responsibility of the multi-agent structural search unit is limited to proposing candidate structures. It is not directly responsible for determining numerical parameters in the structures, nor does it directly judge the quality of candidate models. To facilitate subsequent unified numerical testing and program invocation, this paper requires exploratory agents to directly output executable candidate code. For any candidate structure $h_{i,j}^{(r,\tau)}$, its compilation check result is denoted as
\begin{equation*}
\eta_{i,j}^{(r,\tau)}=\mathrm{CompileCheck}\!\left(h_{i,j}^{(r,\tau)}\right),
\tag{13}
\end{equation*}
where $\eta_{i,j}^{(r,\tau)}\in\{0,1\}$ is the compilation check flag. When $\eta_{i,j}^{(r,\tau)}=0$, the candidate code does not pass syntax or interface-consistency checking and is therefore directly discarded. When $\eta_{i,j}^{(r,\tau)}=1$, the candidate code can enter the subsequent numerical evaluation process.

\subsection{Parameter Determination and Candidate Model Memory Maintenance Mechanism}

Candidate code generated by the multi-agent structural search unit and passing compilation checks essentially provides only candidate model structural skeletons, while its internal undetermined coefficients are still retained as parameter placeholders. To make different candidate structures comparable under a unified standard, their internal parameters must be numerically determined using observational data, and whether a candidate structure should be retained in the candidate model memory must be judged according to its fitting result after parameter determination. Therefore, parameter determination and candidate model memory maintenance are not two isolated steps, but a continuous process in which a candidate structure is converted from ``executable code'' into a ``comparable model.'' The former completes the numerical information in the structure, and the latter completes the retention of preferred candidates and the accumulation of historical experience on this basis.

For any candidate structure $h_{i,j}^{(r,\tau)}$ that passes compilation checking, let its undetermined parameter vector be $\boldsymbol{p}_{i,j}^{(r,\tau)}$. Given dataset $\mathcal{D}$ and current variable space $\mathcal{V}^{(r,\tau)}$, the parameter determination problem can be uniformly represented as
\begin{equation*}
\boldsymbol{p}_{i,j}^{*(r,\tau)}
=
\arg\min_{\boldsymbol{p}}
L_{i,j}^{(r,\tau)}(\boldsymbol{p}),
\tag{14}
\end{equation*}
where $\boldsymbol{p}_{i,j}^{*(r,\tau)}$ is the optimal parameter estimation result for candidate structure $h_{i,j}^{(r,\tau)}$, and $L_{i,j}^{(r,\tau)}(\boldsymbol{p})$ is the fitting loss function corresponding to the current stage. Consistent with the processing approach of LLM-DMD, this paper adopts the mean squared error as the basic objective function for parameter determination and updates parameter placeholders iteratively through the Adam optimizer. Specifically, in the $\tau=f$ stage, the candidate structure fits the true state derivative, and its loss function is written as
\begin{equation*}
L_{i,j}^{(r,f)}(\boldsymbol{p})
=
\frac{1}{n_s}
\sum_{k\in\mathcal{D}}
\left\|
h_{i,j}^{(r,f)}
\!\left(
\mathbf{x}_k,
\hat{\mathbf{y}}_k^{(r,f)},
\hat{\mathbf{u}}_k^{(r,f)},
\boldsymbol{p}
\right)
-
\dot{\mathbf{x}}_k
\right\|_2^2,
\tag{15}
\end{equation*}
where $n_s$ is the total number of sample points, $\dot{\mathbf{x}}_k$ is the true state derivative corresponding to the $k$-th sampling instant, and $\hat{\mathbf{y}}_k^{(r,f)}$ and $\hat{\mathbf{u}}_k^{(r,f)}$ denote the algebraic variables and inputs currently available in the $f$ discovery process, respectively. Correspondingly, in the $\tau=g$ stage, the candidate structure fits the target algebraic quantity to be closed, and its loss function can be written as
\begin{equation*}
L_{i,j}^{(r,g)}(\boldsymbol{p})
=
\frac{1}{n_s}
\sum_{k\in\mathcal{D}}
\left\|
h_{i,j}^{(r,g)}
\!\left(
\mathbf{x}_k,
\hat{\mathbf{y}}_k^{(r,g)},
\hat{\mathbf{u}}_k^{(r,g)},
\boldsymbol{p}
\right)
-
\tilde{\mathbf{y}}_k^{(r)}
\right\|_2^2,
\tag{16}
\end{equation*}
where $\tilde{\mathbf{y}}_k^{(r)}$ denotes the target algebraic quantity identified by the $f$ discovery process in the current round and requiring constraint characterization in the $g$ closure process. Thus, exploratory agents are responsible for generating structural skeletons, while the numerical optimization module is responsible for determining parameter values, and the functional division between the two remains clear.

After obtaining the optimal parameters, the candidate structure $h_{i,j}^{(r,\tau)}$ can be further converted into a candidate model whose parameters have been tuned:
\begin{equation*}
\tilde{h}_{i,j}^{(r,\tau)}
=
\left(
h_{i,j}^{(r,\tau)},
\boldsymbol{p}_{i,j}^{*(r,\tau)}
\right).
\tag{17}
\end{equation*}
To rank different candidate models uniformly, this paper defines their evaluation score as
\begin{equation*}
s_{i,j}^{(r,\tau)}
=
-L_{i,j}^{(r,\tau)}\!\left(\boldsymbol{p}_{i,j}^{*(r,\tau)}\right).
\tag{18}
\end{equation*}
That is, the negative fitting loss under the optimal parameters is used as the scoring metric. Therefore, a higher score indicates that the candidate model fits the observational data better under the current stage. Correspondingly, the objects stored in the candidate model memory are no longer structural templates without parameter tuning, but complete candidate records containing the structural form, optimal parameters, and evaluation score.

Let the candidate model memory of the $i$-th exploratory agent in iteration round $r$ and stage $\tau$ be
\begin{equation*}
B_i^{(r,\tau)}
=
\left\{
b_{i,k}^{(r,\tau)}
\right\}_{k=1}^{\mu},
\tag{19}
\end{equation*}
where $\mu$ denotes the memory capacity and $b_{i,k}^{(r,\tau)}$ denotes the $k$-th candidate record. For each record, this paper retains at least three types of information: the candidate structure $h$, the corresponding optimal parameter $\boldsymbol{p}^*$, and the evaluation score $s$. When error analysis and subsequent prompt construction are required, pointwise residuals and related diagnostic information can also be additionally recorded. The fixed-capacity memory design can prevent high-quality candidates from being forgotten during iteration and can also provide stable historically preferred samples for subsequent prompt construction.

To balance the continuous retention of historically preferred candidates and the dynamic introduction of new candidate models, this paper adopts the classic $(\mu+\lambda)$ elitist retention strategy to update candidate model memories. Suppose the set of candidate models newly generated and parameter-tuned by the $i$-th agent in iteration round $r$ and stage $\tau$ is
\begin{equation*}
\widetilde{\mathcal{H}}_i^{(r,\tau)}
=
\left\{
\left(
h_{i,j}^{(r,\tau)},
\boldsymbol{p}_{i,j}^{*(r,\tau)},
s_{i,j}^{(r,\tau)}
\right)
\right\}_{j=1}^{\lambda},
\tag{20}
\end{equation*}
where $\lambda$ denotes the number of new candidates produced in the current round. Then the memory update process can be written as
\begin{equation*}
B_i^{(r+1,\tau)}
=
\operatorname{Top}_{\mu}
\left(
B_i^{(r,\tau)}
\cup
\widetilde{\mathcal{H}}_i^{(r,\tau)}
\right),
\tag{21}
\end{equation*}
where $\operatorname{Top}_{\mu}(\cdot)$ denotes selecting the top $\mu$ candidate records in descending order of score. The meaning of this update rule is that historical candidate models in the memory and newly generated candidate models in the current round are merged, and only the $\mu$ highest-scoring candidates are retained for the next round. Therefore, the update of the candidate model memory does not depend on subjective rules or manual screening, but is completely built on unified numerical evaluation after parameter tuning.

It should be further emphasized that the candidate model memory not only performs historical retention in this paper, but is also an important context source for subsequent multi-agent search. For the $i$-th exploratory agent, the high-quality candidate records retained in memory $B_i^{(r,\tau)}$ are reinjected into the next-round prompt construction process as few-shot examples, guiding the large language model to continue directed search near existing preferred structures. Because each exploratory agent maintains its own independent candidate model memory, different agents can continuously evolve along different structural paths under a unified evaluation standard, thereby ensuring local search continuity while maintaining structural diversity in the global search process.

\subsection{Coordinator Agent and Its Global Scheduling Responsibilities}

In the proposed multi-agent collaborative discovery framework, coordinator agent $\Omega$ does not directly participate in candidate equation generation or parameter determination. Instead, it is responsible for cross-agent induction of the stage-wise search results of all exploratory agents, and then for generating broadcast summaries and scheduling the global process. Different from exploratory agents, which focus on local structural search, the coordinator agent extracts globally meaningful common information from multiple parallel search chains, so that subsequent search can reduce inefficient repeated exploration while maintaining structural diversity.

In iteration round $r$ and stage $\tau$, the input of the coordinator agent comes from the current candidate model memories of all exploratory agents,
\begin{equation*}
\left\{
B_i^{(r,\tau)}
\right\}_{i=1}^{N_a}.
\tag{22}
\end{equation*}
For the $i$-th exploratory agent, the coordinator first extracts the candidate record with the highest current score, denoted as
\begin{equation*}
b_{i,\mathrm{best}}^{(r,\tau)}
=
\arg\max_{b\in B_i^{(r,\tau)}} s(b),
\qquad i=1,\dots,N_a,
\tag{23}
\end{equation*}
and further defines the global best candidate in the current round and stage $\tau$ as
\begin{equation*}
b_{\mathrm{best}}^{(r,\tau)}
=
\arg\max_{1\le i\le N_a}
s\!\left(
b_{i,\mathrm{best}}^{(r,\tau)}
\right).
\tag{24}
\end{equation*}
On this basis, the coordinator is not limited to comparing candidate scores. Instead, it further combines variable usage, structural patterns, and fitting residuals appearing in high-scoring candidates to analyze the overall search state of the current stage across agents.

Based on the above inputs, when the fixed broadcast period condition is satisfied, the coordinator agent generates a global broadcast summary shared by all exploratory agents, denoted as
\begin{equation*}
\mathcal{C}^{(r,\tau)}
=
\left(
\mathcal{P}^{(r,\tau)},
\Delta\mathcal{Z}^{(r,\tau)},
\mathcal{G}^{(r,\tau)}
\right)
=
\Omega\!\left(
\left\{
B_i^{(r,\tau)}
\right\}_{i=1}^{N_a}
\right),
\tag{25}
\end{equation*}
where $\mathcal{P}^{(r,\tau)}$ denotes a summary of common structural patterns induced from current high-scoring candidates, $\Delta\mathcal{Z}^{(r,\tau)}$ denotes the set of potential missing-variable suggestions identified by the coordinator, and $\mathcal{G}^{(r,\tau)}$ denotes global guidance information for the next round of search. Specifically, $\mathcal{P}^{(r,\tau)}$ is used to extract effective structural features that repeatedly appear in current high-scoring candidates, $\Delta\mathcal{Z}^{(r,\tau)}$ is used to represent variable requirements that have not yet been explicitly included in the existing variable space but may be critical to model closure, and $\mathcal{G}^{(r,\tau)}$ is used to compress common search experience exposed by different search chains to guide the prompt construction of subsequent exploratory agents.

To further clarify the basis on which the coordinator agent identifies potential missing variables and triggers variable extension, this paper adopts a judgment method based on cross-agent consensus. Specifically, the coordinator first extracts the implicit variable requirement information from the current highest-scoring candidate record of each exploratory agent. For the current best candidate $b_{i,\mathrm{best}}^{(r,\tau)}$ of the $i$-th exploratory agent in iteration round $r$ and stage $\tau$, its corresponding local variable suggestion set is denoted as
\begin{equation*}
\Delta\mathcal{Z}_i^{(r,\tau)}
=
\mathrm{VarSuggest}
\!\left(
b_{i,\mathrm{best}}^{(r,\tau)}
\right),
\qquad i=1,\dots,N_a,
\tag{26}
\end{equation*}
where $\mathrm{VarSuggest}(\cdot)$ denotes the variable suggestion extraction operator obtained by the coordinator after inducing the current best candidate according to intermediate quantities that repeatedly appear in candidate structures but have not yet been explicitly included in the current variable space, unclosed dependencies exposed by candidate residuals, and potential physical quantity requirements involved in candidate diagnostic information. Thus, $\Delta\mathcal{Z}_i^{(r,\tau)}$ reflects the local variable extension direction implicitly exposed by the $i$-th exploratory agent through its best candidate in the current round.

After obtaining the local variable suggestions of all exploratory agents, the coordinator further counts the occurrence frequency of different variable requirements in the agent set, and constructs the cross-agent consensus variable set accordingly:
\begin{equation*}
\Delta\mathcal{Z}_{\mathrm{cons}}^{(r,\tau)}
=
\left\{
z \in \bigcup_{i=1}^{N_a}\Delta\mathcal{Z}_i^{(r,\tau)}
\;\middle|\;
\sum_{i=1}^{N_a}
\mathbf{1}\!\left(
z\in\Delta\mathcal{Z}_i^{(r,\tau)}
\right)
\ge m
\right\},
\tag{27}
\end{equation*}
where $m$ denotes the minimum number of agents required to form cross-agent consensus, and $\mathbf{1}(\cdot)$ is the indicator function. When a certain variable suggestion appears simultaneously in the current best candidates of at least $m$ exploratory agents, the coordinator considers that the variable requirement is no longer an accidental guess in a single search chain, but a potential missing-variable signal jointly exposed by multiple heterogeneous search trajectories.

Considering that single-round consensus may still be affected by random generation fluctuations, this paper further requires such consensus variable requirements to remain stable over several consecutive rounds. Let the persistence window length be $K$. If there exists a variable $z$ satisfying
\begin{equation*}
z\in
\bigcap_{\ell=r-K+1}^{r}
\Delta\mathcal{Z}_{\mathrm{cons}}^{(\ell,\tau)},
\tag{28}
\end{equation*}
then the variable requirement is considered to remain valid during the most recent $K$ iterations. If, in combination with the current-stage global best candidate score, the search improvement is found to have slowed significantly, namely
\begin{equation*}
s\!\left(
b_{\mathrm{best}}^{(r,\tau)}
\right)
-
s\!\left(
b_{\mathrm{best}}^{(r-K,\tau)}
\right)
<\varepsilon,
\tag{29}
\end{equation*}
where $\varepsilon>0$ is the threshold used to characterize the degree of search stagnation, then the coordinator determines that the existing variable space is insufficient to support continued significant structural improvement in the current stage, and formally triggers variable extension. Correspondingly, the global variable extension suggestion in the current round can be written as
\begin{equation*}
\Delta\mathcal{Z}^{(r,\tau)}
=
\left(
\bigcap_{\ell=r-K+1}^{r}
\Delta\mathcal{Z}_{\mathrm{cons}}^{(\ell,\tau)}
\right),
\tag{30}
\end{equation*}
and the variable space used by the subsequent stage or next round of search is updated accordingly.

This paper adopts broadcast summaries rather than direct forwarding of raw candidate code. The main consideration is that directly sharing complete candidate code among sub-agents can easily make multiple search chains quickly converge to similar structural neighborhoods, thereby weakening the structural diversity brought by parallel search of heterogeneous exploratory agents. By contrast, a broadcast summary shares only high-level information after cross-agent induction and does not directly expose specific candidate implementations. It can therefore achieve necessary information collaboration while maintaining the relative independence of different exploratory agents on their local search paths.

After generating the global broadcast summary, the coordinator agent feeds it back to all exploratory agents as additional context for prompt construction in subsequent rounds. Correspondingly, for rounds satisfying the broadcast condition, the input of the $i$-th exploratory agent in the next round and stage $\tau$ can be updated as
\begin{equation*}
\mathcal{I}_i^{(r+1,\tau)}
=
\left(
\mathcal{D},
\mathcal{V}^{(r,\tau)},
B_i^{(r,\tau)},
\Pi_i^{(\tau)},
\mathcal{C}^{(r,\tau)}
\right),
\qquad i=1,\dots,N_a.
\tag{31}
\end{equation*}
In rounds that do not trigger broadcasting, each exploratory agent continues to use the most recent broadcast summary as shared context. This indicates that the coordinator agent does not replace exploratory agents in executing candidate generation, but regulates the context on which subsequent search of each sub-agent depends through fixed-period updates of the broadcast summary, allowing agents to continue structural exploration on the basis of shared global cognition.

In addition to cross-agent information induction, the coordinator agent also undertakes global process scheduling. Specifically, the coordinator judges whether the existing variable space can sufficiently support structural search in the current stage according to the current-stage broadcast summary $\mathcal{C}^{(r,\tau)}$ and the global best candidate $b_{\mathrm{best}}^{(r,\tau)}$. When the cross-agent consensus variable set appears stably over several consecutive rounds and the score improvement of the current-stage global best candidate tends to stagnate, the coordinator triggers variable extension and passes $\Delta\mathcal{Z}^{(r,\tau)}$ to subsequent processes. If no significant missing-variable signal appears in the current stage, the framework continues to advance search within the existing variable space. Thus, the coordinator agent simultaneously undertakes two types of responsibilities in the proposed framework, namely cross-agent induction and global process scheduling, so that the multi-agent search process develops from mutually independent parallel exploration into a unified discovery system with controlled information collaboration.

\section{Simulation Verification}
\subsection{Baseline Methods and Evaluation Metrics}

To clearly illustrate the differences between the proposed method and the comparison methods, Table~I summarizes each method from three perspectives: method type, completeness of key-variable priors, and accuracy of predefined function-library priors. MA-LLM-DMD and LLM-DMD both conduct dynamic model discovery under weak prior conditions, while different variants of SINDy and LLM-SR correspond to experimental settings in which key-variable priors or function-library priors change.

\begin{table}[H]
\centering
\caption{Prior information settings of all comparison methods}
\begingroup\small
\begin{tabularx}{\linewidth}{@{}p{0.16\linewidth}p{0.36\linewidth}>{\centering\arraybackslash}p{0.20\linewidth}>{\centering\arraybackslash}p{0.20\linewidth}@{}}
\toprule
Method & Method type & Completeness of key-variable priors & Accuracy of predefined function-library priors \\
\midrule
MA-LLM-DMD & Multi-agent LLM dynamic model discovery method & $\times$ & $\times$ \\
LLM-DMD & Single-agent LLM dynamic model discovery method & $\times$ & $\times$ \\
LLM-SR-AP & Single-equation LLM symbolic regression method & $\checkmark$ & $\times$ \\
LLM-SR-MV & Single-equation LLM symbolic regression method & $\times$ & $\times$ \\
SINDy-AP & Sparse symbolic regression method & $\checkmark$ & $\checkmark$ \\
SINDy-OL & Sparse symbolic regression method & $\checkmark$ & $\times$ \\
SINDy-MV & Sparse symbolic regression method & $\times$ & $\checkmark$ \\
\bottomrule
\end{tabularx}
\endgroup
\end{table}

\noindent Note: AP denotes the accurate-prior setting, OL denotes the overcomplete-library setting, and MV denotes the missing-key-variable setting.

Considering that the experiments in this paper cover both synchronous generators and grid-forming inverters, unified evaluation dimensions are adopted for the two cases to ensure comparability of the subsequent results. Specifically, this paper evaluates different methods from three aspects: time-domain reconstruction and generalization capability, data efficiency and noise robustness, and system-level integration and application capability. Time-domain reconstruction and generalization capability mainly examine the fitting accuracy of the discovered model for key state trajectories and output responses, as well as its extrapolation capability under out-of-distribution conditions. Data efficiency and noise robustness mainly analyze performance trends when training samples are limited and measurement data are contaminated by noise. System-level integration and application capability further examine the dynamic reproduction capability and engineering analysis potential of the discovered model after it is embedded into the system. Based on the above unified comparison settings, this paper verifies the applicability and effectiveness of the proposed method on conventional electromechanical dynamic objects and power-electronic-dominated dynamic objects in the following two cases.

\subsection{Synchronous Generator Dynamic Model Discovery Verification}
\subsubsection{Test System and Dataset Construction}

This paper constructs the IEEE 39-bus 10-machine system in the PSAT dynamic simulation environment, and uses the complete differential-algebraic dynamic process of one synchronous generator unit as the model discovery object. To generate transient data for structure discovery, a three-phase short-circuit disturbance is applied under the system steady-state operating condition. The training condition uses a fault at bus 15, and the testing condition uses a fault at bus 28. Each condition executes a 10~s time-domain simulation with a simulation step of 0.01~s. Based on the above disturbance process, time-series responses of state variables, inputs, and measurable outputs related to the target synchronous generator are further extracted, and a dataset for dynamic model discovery is constructed accordingly.

To be closer to practical measurement conditions, this paper superimposes Gaussian noise with a standard deviation of 1\% of the signal amplitude on the training-set state trajectories to simulate measurement errors. The test set is retained as a fault scenario not involved in training, and is used to test the dynamic reconstruction capability and generalization performance of the discovered model under out-of-distribution disturbances. The ground-truth model of this case adopts the fifth-order Type-I synchronous generator model implemented in PSAT \cite{ref20,ref21}.

\subsubsection{Dynamic Model Discovery Results and Response Verification}

This section analyzes the dynamic model discovery results and response reproduction capability for the synchronous generator case, focusing on model accuracy, search efficiency, and dynamic response consistency of different methods under unseen disturbance scenarios. To this end, this paper uses discovery-process figures, time-domain reconstruction figures, and two result tables to uniformly compare the model discovery quality, generalization performance, and operating cost of the methods. Table~II reports reconstruction accuracy metrics under in-distribution (ID) and out-of-distribution (OOD) conditions, while Table~III reports the iteration rounds, candidate executability rate, and time cost of different stages in the model discovery process.

For quantitative evaluation, this paper uses MAPE and $R^2$ as accuracy metrics to measure the reconstruction capability and generalization performance of the models discovered by different methods for the dynamic responses of the target synchronous generator. It further records the total discovery time $T_{\mathrm{tot}}$, average iteration rounds $N_{\mathrm{it}}$, average compilation pass rate $\rho_{\mathrm{cmp}}$, cumulative LLM call time $T_{\mathrm{LLM}}$, and cumulative parameter optimization time $T_{\mathrm{opt}}$ to compare the search efficiency and computational cost of different methods in the model discovery task. Considering that the LLM-driven discovery process has certain randomness, this paper independently repeats the related stochastic methods 20 times and reports the table results in mean (std) form. For deterministic methods, single-run results are directly given. It should be noted that, for baseline methods that do not involve code generation or explicit parameter optimization, namely SINDy-type methods, the corresponding metrics in the result tables are recorded as ``--''.

\begin{figure}[H]
\centering
\includegraphics[width=0.72\linewidth]{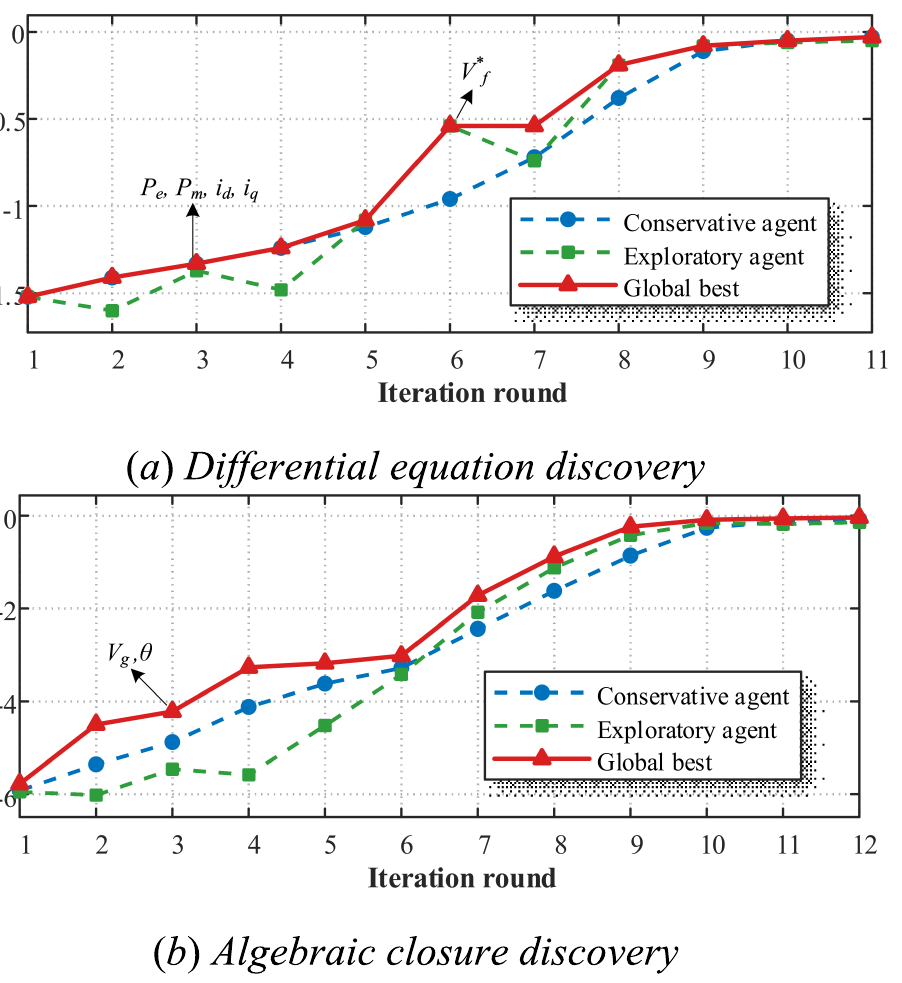}
\caption{Discovery process and convergence characteristics of the synchronous generator dynamic model.}
\end{figure}

\begin{figure}[H]
\centering
\includegraphics[width=0.78\linewidth]{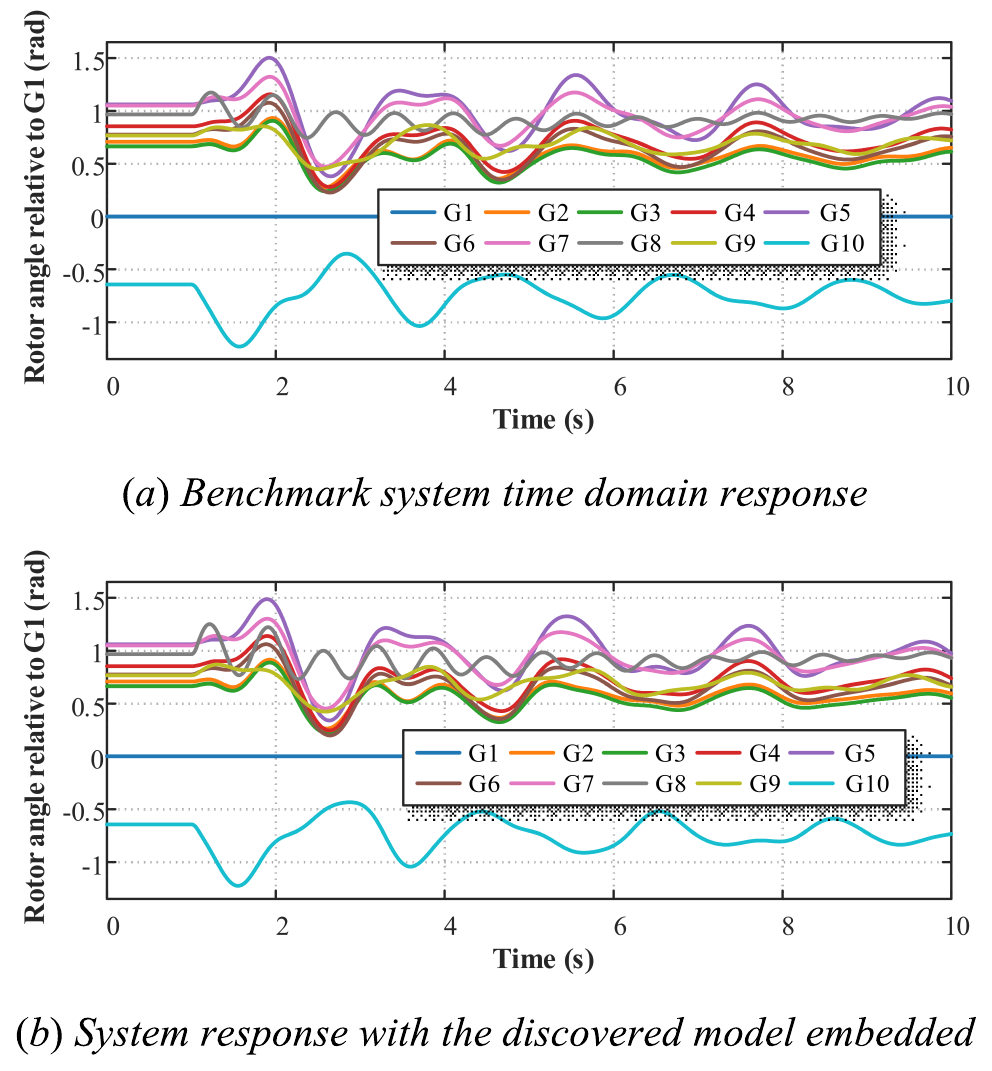}
\caption{Dynamic response reconstruction results of typical states and outputs in the synchronous generator scenario.}
\end{figure}

\begin{table}[H]
\centering
\caption{Accuracy and generalization performance comparison of all methods in the synchronous generator scenario}
\begin{tabularx}{\linewidth}{lCCCC}
\toprule
Method & ID MAPE (\%) & ID $R^2$ & OOD MAPE (\%) & OOD $R^2$ \\
\midrule
MA-LLM-DMD & 0.17 (0.02) & 0.98 (0.01) & 0.19 (0.02) & 0.98 (0.01) \\
LLM-DMD & 0.22 (0.03) & 0.96 (0.03) & 0.21 (0.02) & 0.97 (0.02) \\
LLM-SR-AP & 0.20 (0.03) & 0.97 (0.02) & 0.22 (0.03) & 0.96 (0.03) \\
LLM-SR-MV & 24.85 (10.92) & -7.34 (2.88) & 26.78 (12.78) & -8.28 (3.10) \\
SINDy-AP & 0.62 & 0.97 & 4.72 & 0.59 \\
SINDy-OL & 1.84 & 0.91 & 15.85 & -2.57 \\
SINDy-MV & 16.95 & -5.92 & 17.73 & -6.43 \\
\bottomrule
\end{tabularx}
\end{table}

\begin{table}[H]
\centering
\caption{Discovery efficiency and running-time decomposition of all methods in the synchronous generator scenario}
\begin{tabularx}{\linewidth}{lCCCCC}
\toprule
Method & $T_{\mathrm{tot}}$ (s) & $N_{\mathrm{it}}$ & $\rho_{\mathrm{cmp}}$ (\%) & $T_{\mathrm{LLM}}$ (s) & $T_{\mathrm{opt}}$ (s) \\
\midrule
MA-LLM-DMD & 297 (27) & 16.40 (2.31) & 66.1 (5.4) & 176 (22) & 103 (12) \\
LLM-DMD & 368 (53) & 24.25 (3.85) & 71.5 (6.8) & 228 (41) & 118 (19) \\
LLM-SR-AP & 1495 (205) & 61.35 (8.40) & 76.7 (7.2) & 1168 (176) & 201 (38) \\
LLM-SR-MV & 1511 (160) & 63.80 (9.15) & 72.1 (7.8) & 1182 (168) & 207 (35) \\
SINDy-AP & 38 & -- & -- & -- & -- \\
SINDy-OL & 5200 & -- & -- & -- & -- \\
SINDy-MV & 75 & -- & -- & -- & -- \\
\bottomrule
\end{tabularx}
\end{table}

Table~II shows that MA-LLM-DMD, LLM-DMD, and LLM-SR-AP can all achieve high-accuracy dynamic response reconstruction in the synchronous generator scenario. Their MAPE under both ID and OOD conditions remains at the order of $10^{-1}\%$, and their $R^2$ is stably maintained above 0.96, indicating that these three methods all have strong model expression capability. It is particularly noteworthy that, for methods such as MA-LLM-DMD and LLM-DMD that can recover system dynamic structures relatively completely, performance under the OOD scenario does not show systematic degradation relative to the ID scenario. This indicates that, when the model structure has already been recovered relatively accurately, testing-condition changes are mainly reflected in disturbance forms and numerical response details rather than in structural extrapolation failure. In contrast, LLM-SR-MV and SINDy-MV both show significantly larger errors under ID and OOD scenarios, and their corresponding $R^2$ values are obviously negative, indicating that when key variables are missing, the main source of model error is not insufficient generalization but the inability of the structure itself to close. On the other hand, SINDy-AP and SINDy-OL can rely on strong priors to obtain lower errors in the ID scenario, but their OOD metrics degrade obviously, reflecting that methods based on fixed function libraries and strong priors are more sensitive to prior accuracy and function-library matching, and have relatively limited out-of-distribution stability.

Table~III, together with Fig.~3, further illustrates the effectiveness of the proposed method in organizing the search process. Compared with LLM-DMD, the total discovery time of MA-LLM-DMD decreases from 368 (53)~s to 297 (27)~s, and the average number of iteration rounds decreases from 24.25 (3.85) to 16.40 (2.31), indicating that the multi-agent collaborative framework can complete high-quality candidate model convergence in fewer rounds. Meanwhile, the average compilation pass rate of MA-LLM-DMD is slightly lower than that of LLM-DMD, showing that the proposed framework does not simply shrink the search space, but maintains a balance between exploration and exploitation during search: some agents locally refine existing high-quality candidates, while others explore new expression forms and potential variable relations. Although this mechanism introduces a certain proportion of non-executable candidates, its benefit is that it can expose potential missing variables earlier and form effective structures with interpretability more quickly. The search trajectories shown in Fig.~3 also support this judgment: the multi-agent framework can form effective candidates and identify required variables more quickly in the early stage, show an earlier significant jump after broadcast-summary updating and variable-extension triggering in the middle stage, and enter the stable convergence region with fewer iterations in the later stage.

\subsubsection{In-System Substitution Simulation Verification}

To further verify the system-level embeddability of the discovered model, this paper replaces the original PSAT model of the corresponding unit in the IEEE 39-bus system with the synchronous generator model identified by MA-LLM-DMD, and conducts time-domain simulation again under the same network parameters and control settings. Specifically, the system uses the bus 28 fault in the aforementioned unseen testing scenario as the disturbance condition. By comparing the dynamic responses of the substituted model and the ground-truth model on key states and outputs, the dynamic reproduction capability of the discovered model in an actual system environment is evaluated.

Fig.~4 shows that the synchronous generator model discovered by the proposed method can accurately reproduce the dynamic evolution of key states and outputs under both training conditions and unseen testing conditions, and exhibits high consistency in the amplitude, phase, and damping trend of the main oscillation modes after the fault. This indicates that the high accuracy reflected in Table~II does not remain only at the level of statistical metrics, but can be translated into time-domain dynamic response consistency in in-system substitution simulation. After the discovered model is embedded into the PSAT system, it still remains highly consistent with the ground-truth model, indicating that the proposed method has recovered the dynamic structure of the target object relatively accurately under weak prior conditions. Limited deviations in individual response details mainly come from approximation errors introduced when gradient descent is used for continuous parameter estimation in the parameter optimization stage, rather than from incorrect recognition of the model structure itself. Combining the results of Table~II, Table~III, Fig.~3, and Fig.~4, MA-LLM-DMD can be considered to achieve a good balance among accuracy, generalization, and discovery efficiency in the synchronous generator scenario, and provides a credible model basis for subsequent system-level dynamic simulation and analysis.

\begin{figure}[H]
\centering
\includegraphics[width=0.60\linewidth]{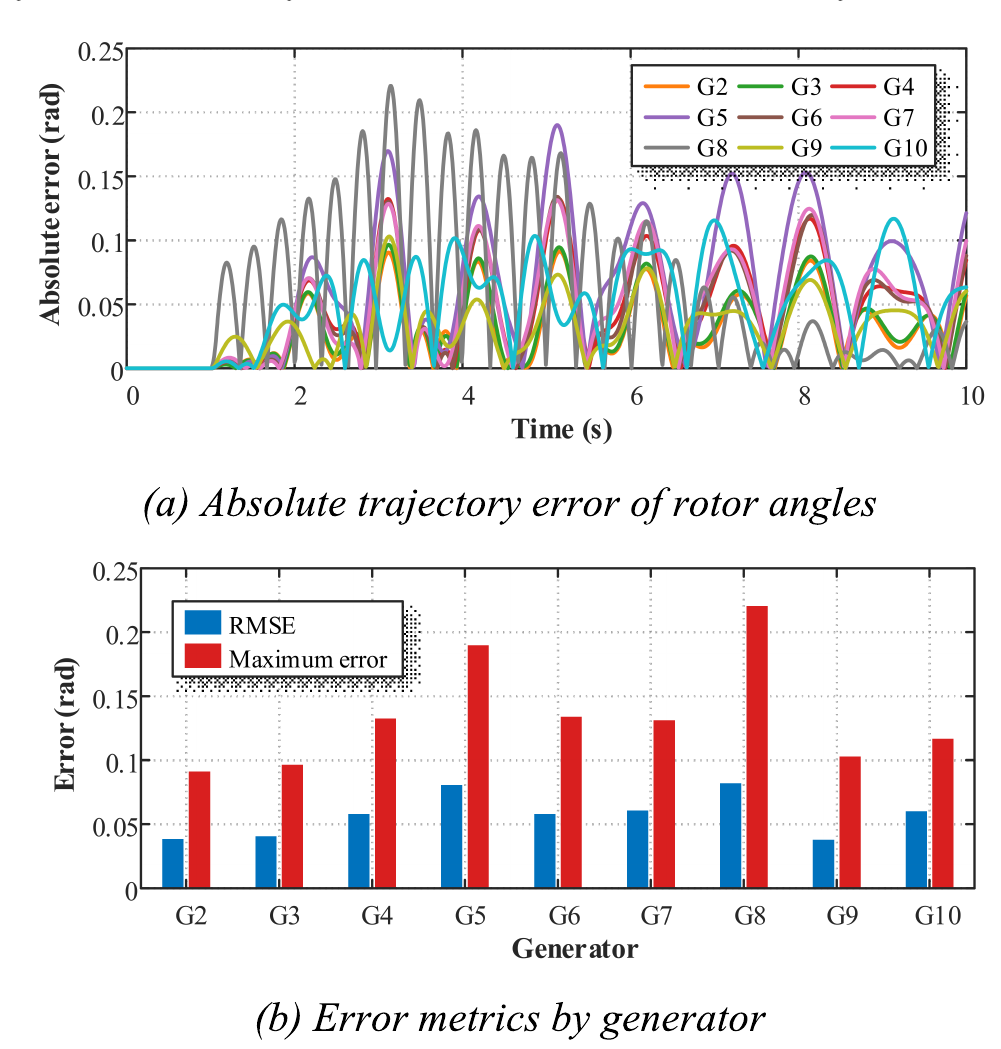}
\caption{Error analysis of system-level substitution simulation in the synchronous generator scenario.}
\end{figure}

Fig.~5 further quantifies the substitution simulation errors of the synchronous generator case. The trajectory error curves remain close to zero during both the fault transient and the post-fault recovery process, indicating that the discovered model does not introduce noticeable dynamic drift after being embedded into the IEEE 39 bus system. The RMSE and maximum absolute error results also show that the main deviations are concentrated around short transient intervals, while the steady state and damping processes are accurately reproduced. This confirms that the proposed method can recover not only offline response trajectories, but also an executable dynamic model with good system simulation consistency.

\subsubsection{Data Efficiency and Noise Robustness Analysis}

This section further evaluates the data utilization efficiency and noise robustness of the proposed method in the synchronous generator scenario from the perspectives of limited training data and contaminated measurement data. For data efficiency, the number of training trajectories used for model discovery can be gradually reduced to compare the accuracy degradation and convergence stability of different methods under limited samples. For noise robustness, stronger Gaussian noise, sparse outliers, and missing samples are further introduced beyond the baseline 1\% Gaussian noise setting, so as to examine the stability of the discovered models under nonideal measurement conditions.

\begin{table}[H]
\centering
\caption{Robustness comparison between MA-LLM-DMD and LLM-DMD under different data contamination conditions}
\begin{tabularx}{\linewidth}{llCCCC}
\toprule
Data contamination type & Contamination level & \multicolumn{2}{c}{MA-LLM-DMD} & \multicolumn{2}{c}{LLM-DMD} \\
\cmidrule(lr){3-4}\cmidrule(lr){5-6}
 & & OOD MAPE (\%) & $T_{\mathrm{tot}}$ (s) & OOD MAPE (\%) & $T_{\mathrm{tot}}$ (s) \\
\midrule
No contamination & 0\% & 0.18 (0.02) & 289 (26) & 0.19 (0.02) & 355 (49) \\
Gaussian noise & 1\% & 0.19 (0.03) & 297 (27) & 0.21 (0.02) & 368 (53) \\
 & 3\% & 0.26 (0.05) & 314 (32) & 0.29 (0.06) & 381 (55) \\
 & 5\% & 0.33 (0.09) & 341 (41) & 0.41 (0.13) & 431 (70) \\
Sparse outliers & 1\% & 0.29 (0.05) & 326 (37) & 0.36 (0.06) & 394 (58) \\
 & 3\% & 0.58 (0.12) & 387 (49) & 0.74 (0.14) & 451 (67) \\
 & 5\% & 0.91 (0.24) & 463 (68) & 1.26 (0.29) & 536 (82) \\
Missing samples & 1\% & 0.21 (0.03) & 304 (30) & 0.23 (0.03) & 372 (54) \\
 & 3\% & 0.30 (0.04) & 323 (35) & 0.33 (0.02) & 389 (57) \\
 & 5\% & 0.43 (0.07) & 359 (45) & 0.57 (0.05) & 423 (63) \\
\bottomrule
\end{tabularx}
\end{table}

As shown in Table~IV, under different types of data contamination, MA-LLM-DMD always exhibits a flatter performance degradation trend than LLM-DMD, indicating that the multi-agent collaborative framework has stronger robustness under nonideal measurement conditions. Taking Gaussian noise as an example, when the noise level increases from 1\% to 5\%, the OOD MAPE of MA-LLM-DMD increases from 0.19\% to 0.33\%, while that of LLM-DMD increases from 0.21\% to 0.41\%. Under sparse outlier conditions, the performance gap between the two methods further widens. When the outlier ratio reaches 5\%, the OOD MAPE of MA-LLM-DMD is 0.91\%, clearly lower than the 1.26\% of LLM-DMD. For missing samples, the proposed method also maintains a lower error level, keeping OOD MAPE at 0.43\% under a 5\% missing rate, while LLM-DMD rises to 0.57\%. Meanwhile, as data contamination becomes stronger, the total discovery time of MA-LLM-DMD increases, but remains lower overall than that of LLM-DMD. For example, under 5\% Gaussian noise, 5\% sparse outliers, and 5\% missing samples, its total discovery times are 341~s, 463~s, and 359~s, respectively, all lower than the 431~s, 536~s, and 423~s of LLM-DMD. These results indicate that the proposed method has better adaptability to nonideal measurement data while maintaining its search efficiency advantage.

This phenomenon is consistent with the mechanism design of the proposed multi-agent framework. First, parallel heterogeneous search can reduce the cumulative effect of misjudgments along a single path. Under noisy conditions, a single search chain is more likely to repeatedly correct around accidental residual patterns, whereas the multi-agent framework allows different inductive biases to coexist in parallel, preventing misleading information caused by local noise from dominating the entire discovery process. Second, candidate memory retention and the coordination broadcast mechanism reinforce stable structures and suppress accidental structures. The information entering the memory and broadcast summary is not a one-time candidate, but high-quality results retained after parameter determination and unified scoring. Therefore, noise-induced transient high-scoring structures usually have difficulty forming persistent consensus among multiple agents and thus have difficulty influencing subsequent search over the long term. In addition, variable extension is not directly triggered by single-round local anomalies, but is built jointly on cross-agent consensus, persistence checking, and global search-stagnation criteria. Therefore, it is less likely to misjudge missing-variable signals because of noise disturbances, and can avoid the amplification effect of erroneous variable extension on subsequent discovery. Overall, the proposed method shows good stability under the three types of nonideal data conditions, namely Gaussian noise, sparse outliers, and missing samples, indicating that the multi-agent collaborative discovery mechanism can effectively mitigate the interference of measurement contamination with dynamic model discovery.

\subsection{Grid-Forming Inverter Dynamic Model Discovery Verification}
\subsubsection{Test System and Dataset Construction}

To evaluate the effectiveness of the proposed multi-agent dynamic model discovery method in grid-connected power-electronic systems, this paper constructs a grid-forming inverter test system based on virtual synchronous generator (VSG) control in the Simulink environment. The test object consists of a VSG-controlled inverter, an LC filter, and an external infinite grid (stiff grid). The inverter output is connected to the point of common coupling (PCC) after passing through the LC filter, and is further connected to the infinite grid. The main parameter configuration of the test system is shown in Table~V. The system includes outer synchronization and voltage-regulation control, cascaded voltage-current control, and transient dynamics jointly determined by the filtering link and grid interaction. It can reflect typical dynamic behavior of grid-forming inverters under disturbance conditions and is therefore suitable as a validation object for the dynamic model discovery task in this paper.

To construct transient data required for model discovery, this paper uses the Simulink test system as the dynamic data generation platform, and excites the system response by adjusting the voltage at the PCC. Specifically, the system first runs at a given steady-state operating point. Then, at a specified instant, a disturbance is applied to the PCC voltage amplitude so that it varies within 1.0--1.8 p.u., causing the system to deviate from the original equilibrium state and exciting the main transient modes in VSG outer synchronization and voltage-regulation links, LC filter dynamics, and grid-connected power exchange. To simulate practical situations in data-driven modeling where training samples are limited and operating conditions have distribution shifts, and to evaluate the model discovery capability of the proposed method under limited training samples, this paper generates eight dynamic response trajectories in total. Six of them are used for model discovery and parameter identification, and the other two are used as out-of-distribution (OOD) test trajectories to examine extrapolation and generalization performance of the discovered model on conditions not involved in training. Based on the above disturbance settings, this paper further extracts time-series data of key electrical quantities and control state variables in each trajectory, and constructs the dataset required for subsequent dynamic model discovery and result verification.

\begin{longtable}{llrl}
\caption{Parameters of the VSG-controlled grid-connected inverter test system}\\
\toprule
Parameter & Symbol & Value & Unit \\
\midrule
\endfirsthead
\caption[]{Parameters of the VSG-controlled grid-connected inverter test system (continued)}\\
\toprule
Parameter & Symbol & Value & Unit \\
\midrule
\endhead
\multicolumn{4}{l}{\textit{System rated parameters}}\\
DC-side voltage & $V_{in}$ & 800 & V \\
Carrier amplitude & $V_{tri}$ & 400 & V \\
Rated voltage amplitude & $V_g$ & 220 & V \\
Active power reference & $P^{\star}$ & 100 & kW \\
Reactive power reference & $Q^{\star}$ & 0 & var \\
Rated frequency & $f_{line}$ & 50 & Hz \\
Switching frequency & $f_{sw}$ & 10 & kHz \\
Sampling frequency & $f_{sample}$ & 20 & kHz \\
Base angular frequency & $\omega_n$ & 314.16 & rad/s \\
\multicolumn{4}{l}{\textit{Filter parameters}}\\
Inverter-side inductance & $L_1$ & 3.2 & mH \\
Filter capacitance & $C$ & 50 & $\mu$F \\
\multicolumn{4}{l}{\textit{VSG control parameters}}\\
Active-frequency regulation coefficient & $D_p$ & 5300 & -- \\
Virtual inertia & $J$ & 0.081 & -- \\
Reactive-voltage regulation coefficient & $D_q$ & 3210 & -- \\
Reactive integral regulation coefficient & $K$ & 7.1 & -- \\
Reactive proportional regulation coefficient & $K_{p,Q}$ & 0 & -- \\
Virtual resistance & $R_v$ & 1.1 & $\Omega$ \\
Virtual inductance & $L_v$ & 5.12 & mH \\
Voltage-loop proportional coefficient & $K_{p,Vreg}$ & 0.4 & -- \\
Voltage-loop integral coefficient & $K_{i,Vreg}$ & 90 & -- \\
Current-loop proportional coefficient & $K_{p,Ireg}$ & 11 & -- \\
Current-loop integral coefficient & $K_{i,Ireg}$ & 0 & -- \\
\bottomrule
\end{longtable}

\subsubsection{Ground-Truth Dynamic Model}

To clearly explain the true generation mechanism of the experimental data in the test system, this paper further gives the ground-truth dynamic model corresponding to the Simulink test system. This model is used only to explain the source of simulation data generation and to provide a unified reference for variable definitions and interpretations in subsequent experiments. Considering that the test system is essentially composed of the physical object, outer synchronization control, and cascaded voltage-current inner loops, this paper no longer compresses it uniformly into a single black-box mapping, but represents it stage by stage according to control hierarchy as a nonlinear closed-loop dynamic system described in a synchronous rotating $dq$ coordinate system.

\paragraph{1) LC filter dynamics.}
For the test system used in this paper, the physical state vector is defined as
\begin{equation*}
x_{\mathrm{phy}}
=
\begin{bmatrix}
i_{fd} & i_{fq} & v_{cd} & v_{cq}
\end{bmatrix}^{\top}.
\tag{32}
\end{equation*}
Here, $i_{fd}$ and $i_{fq}$ denote the components of filter inductor current on the $d$ and $q$ axes, respectively, while $v_{cd}$ and $v_{cq}$ denote the components of filter capacitor voltage on the $d$ and $q$ axes, respectively. Correspondingly, the physical dynamics on the filter side can be written as
\begin{equation*}
L_1 \dot{i}_{fd}=u_d-v_{cd}+\omega L_1 i_{fq},
\tag{33}
\end{equation*}
\begin{equation*}
L_1 \dot{i}_{fq}=u_q-v_{cq}-\omega L_1 i_{fd},
\tag{34}
\end{equation*}
\begin{equation*}
C \dot{v}_{cd}=i_{fd}-i_{od}-i_{Ld}+\omega C v_{cq},
\tag{35}
\end{equation*}
\begin{equation*}
C \dot{v}_{cq}=i_{fq}-i_{oq}-i_{Lq}-\omega C v_{cd}.
\tag{36}
\end{equation*}
Here, $u_d$ and $u_q$ are the components of inverter bridge output voltage in the $dq$ coordinate system, $i_{od}$ and $i_{oq}$ are the output current components flowing to the external network, and $i_{Ld}$ and $i_{Lq}$ are the local load current components. The above equations describe the basic electromagnetic transient process jointly determined by the filter inductor and filter capacitor, and form the physical basis of the dynamic evolution of the entire test system.

\paragraph{2) Power calculation and outer VSG control.}
At the control level, the test system adopts a hierarchical control structure consisting of the outer VSG control, cascaded voltage loop, and current loop. The role of the outer VSG control is to generate internal frequency and voltage references according to active and reactive power deviations, thereby characterizing the synchronization and voltage-regulation behavior of the grid-forming inverter. Define the outer control state as
\begin{equation*}
x_{\mathrm{vsg}}
=
\begin{bmatrix}
\theta & \omega & \xi_Q
\end{bmatrix}^{\top},
\end{equation*}
where $\theta$ and $\omega$ denote the internal angle and angular frequency of the VSG, respectively, and $\xi_Q$ denotes the integral state of the reactive voltage regulator.

The output active power and reactive power are given by the following algebraic relations:
\begin{equation*}
P=v_{cd}i_{od}+v_{cq}i_{oq},
\tag{37}
\end{equation*}
\begin{equation*}
Q=v_{cq}i_{od}-v_{cd}i_{oq}.
\tag{38}
\end{equation*}
On this basis, the outer VSG dynamics can be represented as
\begin{equation*}
\dot{\theta}=\omega-\omega_n,
\tag{39}
\end{equation*}
\begin{equation*}
J\dot{\omega}=P^{\star}-P-D_p(\omega-\omega_n),
\tag{40}
\end{equation*}
\begin{equation*}
\dot{\xi}_Q=Q^{\star}-Q,
\tag{41}
\end{equation*}
where $\omega_n$ is the rated angular frequency, $J$ is the virtual inertia, $D_p$ is the active-frequency regulation coefficient, and $P^\star$ and $Q^\star$ are the active and reactive power references, respectively.

Based on the reactive power deviation, the outer control further generates the voltage amplitude reference
\begin{equation*}
V^{\star}=V_g+D_q(Q^{\star}-Q)+K_{i,Q}\xi_Q+K_{p,Q}(Q^{\star}-Q),
\tag{42}
\end{equation*}
and takes
\begin{equation*}
v_{cd}^{\star}=V^{\star}, \qquad v_{cq}^{\star}=0.
\tag{43}
\end{equation*}
Here, $V_g$ denotes the rated voltage amplitude, $D_q$ is the reactive-voltage regulation coefficient, and $K_{p,Q}$ and $K_{i,Q}$ are the proportional and integral parameters of the reactive voltage regulator, respectively. The above setting indicates that, in the synchronous rotating coordinate system adopted here, the capacitor voltage reference vector is aligned with the $d$ axis, so its $q$-axis reference component is zero.

\paragraph{3) Voltage-loop control.}
After obtaining the voltage reference from the outer loop, the voltage loop further generates the filter inductor current reference according to capacitor voltage deviations. Define the voltage-loop integral state as
\begin{equation*}
x_{\mathrm{v}}
=
\begin{bmatrix}
\xi_{vd} & \xi_{vq}
\end{bmatrix}^{\top},
\end{equation*}
and its dynamic equations are
\begin{equation*}
\dot{\xi}_{vd}=v_{cd}^{\star}-v_{cd},
\end{equation*}
\begin{equation*}
\dot{\xi}_{vq}=v_{cq}^{\star}-v_{cq}.
\end{equation*}
Correspondingly, the current references output by the voltage loop can be written as
\begin{equation*}
i_{fd}^{\star}
=
K_{p,Vreg}(v_{cd}^{\star}-v_{cd})
+K_{i,Vreg}\xi_{vd}
-\omega C v_{cq}
+i_{od}+i_{Ld},
\tag{44}
\end{equation*}
\begin{equation*}
i_{fq}^{\star}
=
K_{p,Vreg}(v_{cq}^{\star}-v_{cq})
+K_{i,Vreg}\xi_{vq}
+\omega C v_{cd}
+i_{oq}+i_{Lq}.
\tag{45}
\end{equation*}
Here, $K_{p,Vreg}$ and $K_{i,Vreg}$ are the proportional and integral parameters of the voltage-loop PI controller, respectively. It can be seen that, in addition to conventional PI regulation terms, the above expressions also contain cross-coupling compensation terms introduced by the synchronous rotating coordinate transformation and feedforward terms composed of output current and load current, so that the voltage loop can more directly reflect capacitor dynamics and load requirements.

\paragraph{4) Current-loop control.}
The current reference generated by the voltage loop is further input to the current loop to form the inverter bridge output voltage command. Define the current-loop integral state as
\begin{equation*}
x_{\mathrm{i}}
=
\begin{bmatrix}
\xi_{id} & \xi_{iq}
\end{bmatrix}^{\top},
\end{equation*}
and its dynamics satisfy
\begin{equation*}
\dot{\xi}_{id}=i_{fd}^{\star}-i_{fd},
\end{equation*}
\begin{equation*}
\dot{\xi}_{iq}=i_{fq}^{\star}-i_{fq}.
\end{equation*}
The corresponding current-loop control laws are
\begin{equation*}
u_d
=
K_{p,Ireg}(i_{fd}^{\star}-i_{fd})
+K_{i,Ireg}\xi_{id}
-\omega L_1 i_{fq}
+v_{cd}
-R_v i_{od}
+\omega L_v i_{oq},
\tag{46}
\end{equation*}
\begin{equation*}
u_q
=
K_{p,Ireg}(i_{fq}^{\star}-i_{fq})
+K_{i,Ireg}\xi_{iq}
+\omega L_1 i_{fd}
+v_{cq}
-R_v i_{oq}
-\omega L_v i_{od}.
\tag{47}
\end{equation*}
Here, $K_{p,Ireg}$ and $K_{i,Ireg}$ are the proportional and integral parameters of the current-loop PI controller, respectively, while $R_v$ and $L_v$ denote the virtual resistance and virtual inductance coefficients in the virtual impedance. In addition to PI regulation terms, the above control laws explicitly include filter inductor coupling compensation, capacitor voltage feedforward, and virtual impedance terms, jointly determining the inverter bridge output voltages $u_d$ and $u_q$.

To keep the ground truth model consistent with the DAE formulation, the above model can be divided into a differential subsystem and an algebraic closure subsystem. Equations (33)--(36) and (39)--(41) describe the evolution of physical states and outer control states, forming the differential equation $f(\cdot)$. Equations (37)--(38) and (42)--(47) describe power calculation, reference generation, voltage-loop closure, and current-loop closure, forming the algebraic equation $g(\cdot)$.

\paragraph{5) Time-domain reconstruction and generalization capability verification.}
This section verifies the model discovery effect of the proposed method in the grid-forming inverter scenario from three aspects: discovery trajectories, ID/OOD reconstruction accuracy, and operating cost.

\begin{figure}[H]
\centering
\includegraphics[width=0.75\linewidth]{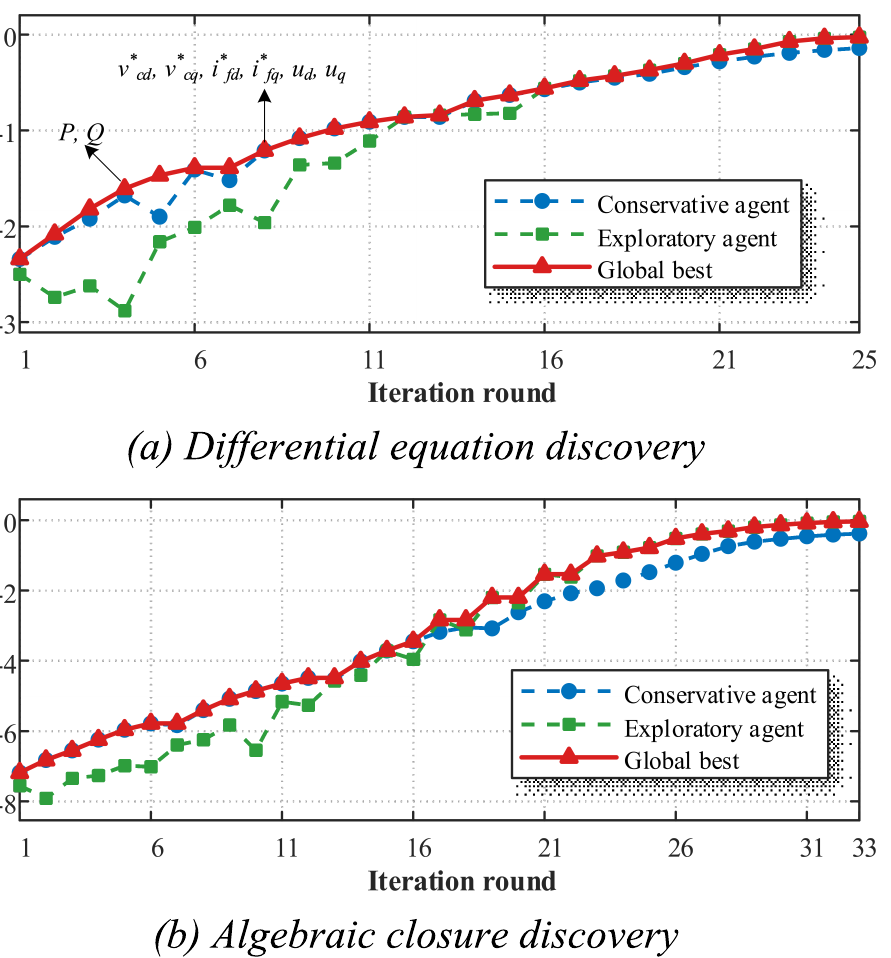}
\caption{Discovery process and convergence characteristics of the grid-forming inverter dynamic model.}
\end{figure}

The DE stage sequentially discovers $\{P,Q\}$, $\{v_{cd}^{\star},v_{cq}^{\star}\}$, $\{i_{fd}^{\star},i_{fq}^{\star}\}$, and $\{u_d,u_q\}$; the AE stage further discovers the algebraic closure relations among the above variables, including power calculation, reference generation, voltage-loop control laws, and current-loop control laws.

As shown in Fig.~6, as key variables and their closure relations are progressively incorporated, the scores of both the DE and AE stages increase in a stepwise manner and eventually approach 0. The jumps in Fig.~6(a) mainly correspond to the identification of intermediate variables such as power variables, reference variables, and bridge output voltages. The jumps in Fig.~6(b) correspond to the progressive closure of algebraic relations among the identified variables.

From the trajectories of different agents, the conservative agent converges more smoothly overall, while the exploratory agent fluctuates more obviously but can produce better candidates in some rounds. The global best curve retains and switches among high-quality candidates from different search paths, and therefore shows obvious improvements in several key rounds.

\begin{table}[H]
\centering
\caption{Accuracy and generalization performance comparison of all methods in the grid-forming inverter scenario}
\begin{tabularx}{\linewidth}{lCCCC}
\toprule
Method & ID MAPE (\%) & ID $R^2$ & OOD MAPE (\%) & OOD $R^2$ \\
\midrule
MA-LLM-DMD & 0.55 (0.05) & 0.96 (0.01) & 0.58 (0.06) & 0.96 (0.01) \\
LLM-DMD & 0.83 (0.36) & 0.92 (0.06) & 0.91 (0.42) & 0.91 (0.07) \\
LLM-SR-AP & 1.12 (0.58) & 0.88 (0.10) & 1.28 (0.71) & 0.85 (0.13) \\
LLM-SR-MV & 18.74 (6.35) & -3.92 (1.44) & 24.61 (8.72) & -6.15 (2.36) \\
SINDy-AP & 2.84 & 0.86 & 8.73 & 0.41 \\
SINDy-OL & 6.92 & 0.52 & 19.38 & -2.84 \\
SINDy-MV & 27.45 & -7.86 & 34.72 & -10.95 \\
\bottomrule
\end{tabularx}
\end{table}

\begin{table}[H]
\centering
\caption{Discovery efficiency and running-time decomposition of all methods in the grid-forming inverter scenario}
\begin{tabularx}{\linewidth}{lCCCCC}
\toprule
Method & $T_{\mathrm{tot}}$ (s) & $N_{\mathrm{it}}$ & $\rho_{\mathrm{cmp}}$ (\%) & $T_{\mathrm{LLM}}$ (s) & $T_{\mathrm{opt}}$ (s) \\
\midrule
MA-LLM-DMD & 626 (74) & 23.85 (3.62) & 58.4 (6.1) & 372 (51) & 214 (33) \\
LLM-DMD & 842 (118) & 35.70 (5.44) & 63.8 (7.3) & 531 (86) & 265 (47) \\
LLM-SR-AP & 2315 (330) & 82.45 (11.60) & 68.5 (8.0) & 1789 (274) & 361 (72) \\
LLM-SR-MV & 2388 (365) & 86.20 (12.35) & 64.9 (8.4) & 1842 (302) & 374 (81) \\
SINDy-AP & 96 & -- & -- & -- & -- \\
SINDy-OL & 7800 & -- & -- & -- & -- \\
SINDy-MV & 184 & -- & -- & -- & -- \\
\bottomrule
\end{tabularx}
\end{table}

The MAPE values of MA-LLM-DMD under ID and OOD scenarios are 0.55\% and 0.58\%, respectively, and the corresponding $R^2$ values are both maintained at 0.96, indicating that it does not show obvious performance degradation under unseen disturbance conditions. For objects such as grid-forming inverters that contain multilayer control closure relations, close ID and OOD errors usually imply that the discovered model has captured the main structural relations, and the remaining error mainly comes from continuous parameter fitting, numerical differentiation, and simulation discretization, rather than structural-level extrapolation failure. Combined with the phenomenon that the global best curve in Fig.~6 approaches 0 stably in the later stage, the multi-agent framework can be considered to have relatively completely recovered the filter dynamics, VSG outer loop, and inner-loop control closure relations.

In contrast, although the average errors of LLM-DMD and LLM-SR-AP remain within an acceptable range, result fluctuations increase obviously. The ID/OOD MAPE values of LLM-DMD are 0.83\% and 0.91\%, and the standard deviations reach 0.36\% and 0.42\%. The OOD MAPE standard deviation of LLM-SR-AP further increases to 0.71\%. This indicates that single-agent search has insufficient stability in the more complex GFM inverter scenario: some runs can recover structures close to the correct ones, while other runs may deviate from the correct search direction at power calculation, reference generation, or current-loop closure relations. In other words, the increase in error is not entirely due to parameter estimation, but more reflects the randomness of single-path structural search and the accumulation of local misjudgments. MA-LLM-DMD reduces the influence of such accidental structural errors through multi-agent parallel exploration and candidate memory retention, and therefore exhibits smaller variance and more stable ID/OOD consistency across multiple independent runs.

When key variables are missing, the OOD MAPE values of LLM-SR-MV and SINDy-MV increase to 24.61\% and 34.72\%, respectively, and their $R^2$ values are both negative, indicating that these methods cannot form closed dynamic models. SINDy-AP has an ID MAPE of 2.84\%, but its OOD MAPE increases to 8.73\%. The OOD MAPE of SINDy-OL further increases to 19.38\%. This shows that sparse regression methods based on fixed function libraries are sensitive to variable priors and function-library matching, and are prone to out-of-distribution performance degradation when disturbance amplitudes change.

The total discovery time of MA-LLM-DMD is 626~s, lower than the 842~s of LLM-DMD. Its average iteration rounds are 23.85, fewer than the 35.70 of LLM-DMD, indicating that parallel exploration and broadcast feedback reduce the number of rounds required for single-path trial and error. Its average compilation pass rate is 58.4\%, lower than the 63.8\% of LLM-DMD, reflecting that the multi-agent framework generates more candidates with different structural forms. Combining total time and final accuracy shows that this candidate diversity, although introducing a certain proportion of non-executable structures, helps discover effective algebraic closure paths earlier. In contrast, the total discovery times of LLM-SR-AP and LLM-SR-MV both exceed 2300~s, indicating that, when the staged discovery mechanism for DAE structures is absent, directly searching in a single-equation expression space significantly increases computational cost.

Combining Fig.~6, Table~VI, and Table~VII, it can be seen that the performance differences in the grid-forming inverter case are not determined only by parameter fitting accuracy, but mainly by whether a method can discover and close control intermediate variables. MA-LLM-DMD shows better comprehensive performance in discovery process, reconstruction accuracy, and search efficiency through multi-agent parallel search, candidate memory retention, and the coordinator broadcast mechanism.

\subsubsection{Data Efficiency and Noise Robustness Analysis}

This section further examines model discovery stability when training samples are limited and measurement data are contaminated. Because the grid-forming inverter contains faster electromagnetic transients and stronger control-algebraic coupling, its discovery process is more sensitive to noise, outliers, and insufficient operating-condition coverage.

For data efficiency analysis, this paper can gradually reduce the number of training trajectories used for model discovery while keeping the testing trajectories unchanged, for example by using 6, 4, and 2 training trajectories, respectively, for discovery experiments, and compare the MAPE, $R^2$, and discovery time of each method on OOD testing trajectories. If the error of MA-LLM-DMD grows more slowly after the number of training trajectories is reduced, and the finally discovered model can still retain the main physical and control structures, this indicates that multi-agent collaborative search can more fully use the dynamic information in limited samples and improve discovery stability under low-sample conditions through candidate memories and broadcast summaries.

For noise robustness analysis, this paper introduces three types of data contamination on the basis of the baseline dataset: Gaussian noise, sparse outliers, and missing samples. Gaussian noise is used to simulate continuous measurement errors, sparse outliers are used to simulate communication anomalies or sampling spikes, and missing samples are used to simulate short-term data loss. Under each contamination scenario, the discovery process, evaluation metrics, and testing conditions remain consistent, and the OOD reconstruction error and total discovery time of MA-LLM-DMD and LLM-DMD are compared with emphasis.

\begin{table}[H]
\centering
\caption{Robustness comparison of grid-forming inverter model discovery under different data contamination conditions}
\begin{tabularx}{\linewidth}{llCCCC}
\toprule
Data contamination type & Contamination level & \multicolumn{2}{c}{MA-LLM-DMD} & \multicolumn{2}{c}{LLM-DMD} \\
\cmidrule(lr){3-4}\cmidrule(lr){5-6}
 & & OOD MAPE (\%) & $T_{\mathrm{tot}}$ (s) & OOD MAPE (\%) & $T_{\mathrm{tot}}$ (s) \\
\midrule
No contamination & 0\% & 0.68 (0.09) & 602 (69) & 0.98 (0.16) & 815 (110) \\
Gaussian noise & 1\% & 0.72 (0.10) & 626 (74) & 1.08 (0.18) & 842 (118) \\
 & 3\% & 0.96 (0.15) & 694 (85) & 1.45 (0.27) & 935 (136) \\
 & 5\% & 1.32 (0.24) & 782 (104) & 2.05 (0.42) & 1088 (174) \\
Sparse outliers & 1\% & 1.05 (0.20) & 711 (92) & 1.67 (0.34) & 972 (151) \\
 & 3\% & 1.84 (0.39) & 893 (126) & 2.91 (0.63) & 1196 (205) \\
 & 5\% & 2.96 (0.71) & 1128 (188) & 4.58 (1.05) & 1515 (278) \\
Missing samples & 1\% & 0.81 (0.12) & 655 (80) & 1.22 (0.22) & 884 (126) \\
 & 3\% & 1.18 (0.18) & 742 (96) & 1.78 (0.33) & 1011 (152) \\
 & 5\% & 1.67 (0.29) & 858 (119) & 2.56 (0.50) & 1254 (218) \\
\bottomrule
\end{tabularx}
\end{table}

Table~VIII shows the degradation trend of model discovery performance as the degree of data contamination changes. Overall, MA-LLM-DMD shows lower OOD error and shorter total discovery time than LLM-DMD under all three types of data contamination. Taking Gaussian noise as an example, when the noise level increases from 1\% to 5\%, the OOD MAPE of MA-LLM-DMD increases from 0.72\% to 1.32\%, while that of LLM-DMD increases from 1.08\% to 2.05\%. Under sparse outlier conditions, the performance gap between the two methods further widens. When the outlier ratio reaches 5\%, the OOD MAPE of MA-LLM-DMD is 2.96\%, clearly lower than the 4.58\% of LLM-DMD. For missing samples, the proposed method also maintains a relatively flat error increase, keeping OOD MAPE at 1.67\% under a 5\% missing rate, while LLM-DMD rises to 2.56\%.

The above results show that the multi-agent collaborative framework can effectively mitigate the misleading influence of noise on the structural search process. The reason is that a single-agent search chain is more likely to over-correct around local residual patterns under noisy data, whereas multi-agent parallel search can retain multiple candidate structural paths. After parameter determination and unified scoring, high-scoring candidates accidentally induced by noise usually have difficulty forming stable consensus among multiple agents, and therefore are less likely to dominate subsequent search over the long term. In addition, when the coordinator triggers variable extension, it relies on cross-agent consensus and search stagnation signals rather than single-round candidate errors. It can therefore reduce the risk of erroneous variable extension caused by outliers or missing samples. Overall, even though the grid-forming inverter scenario is more noise-sensitive than the synchronous generator scenario, the proposed method can still maintain good model discovery stability under different data contamination conditions.

\subsubsection{System-Level Integration and Application Verification}

Beyond single-device trajectory fitting, this section evaluates whether the discovered grid-forming inverter model can be embedded into system-level simulation. The ground-truth control and filter model in the original Simulink system is replaced by the MA-LLM-DMD discovered model, while the external grid, voltage disturbance, and operating parameters remain unchanged. The voltage, current, power, and frequency responses of the substituted model are then compared with those of the ground-truth model to assess its executability and dynamic consistency.

\begin{figure}[H]
\centering
\includegraphics[width=0.78\linewidth]{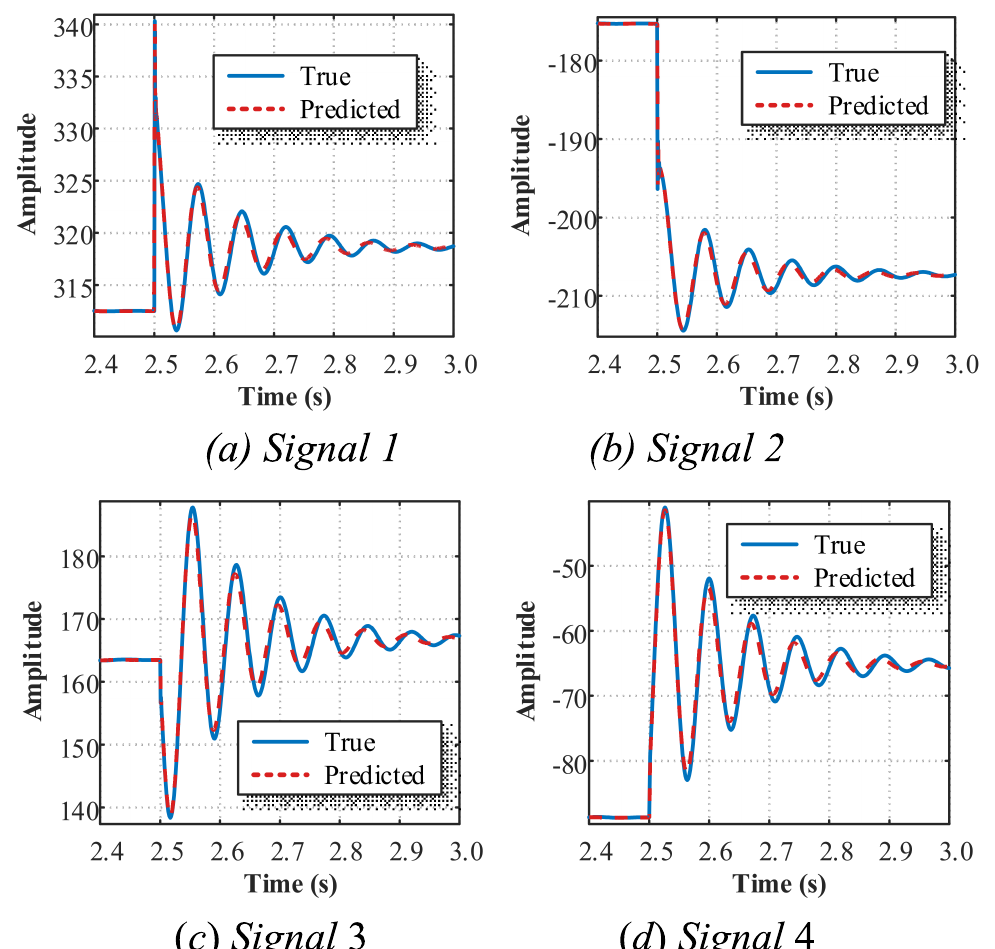}
\caption{Dynamic response results of system-level substitution simulation in the grid-forming inverter scenario.}
\end{figure}

Fig.~7 shows the system level time domain responses after replacing the original Simulink inverter model with the MA-LLM-DMD discovered model. Under unseen PCC voltage disturbances, the substituted model closely follows the ground-truth responses in both fast electromagnetic transients and slower control regulation processes, indicating that the discovered model can be executed stably in the closed-loop Simulink environment.

\begin{figure}[H]
\centering
\includegraphics[width=0.64\linewidth]{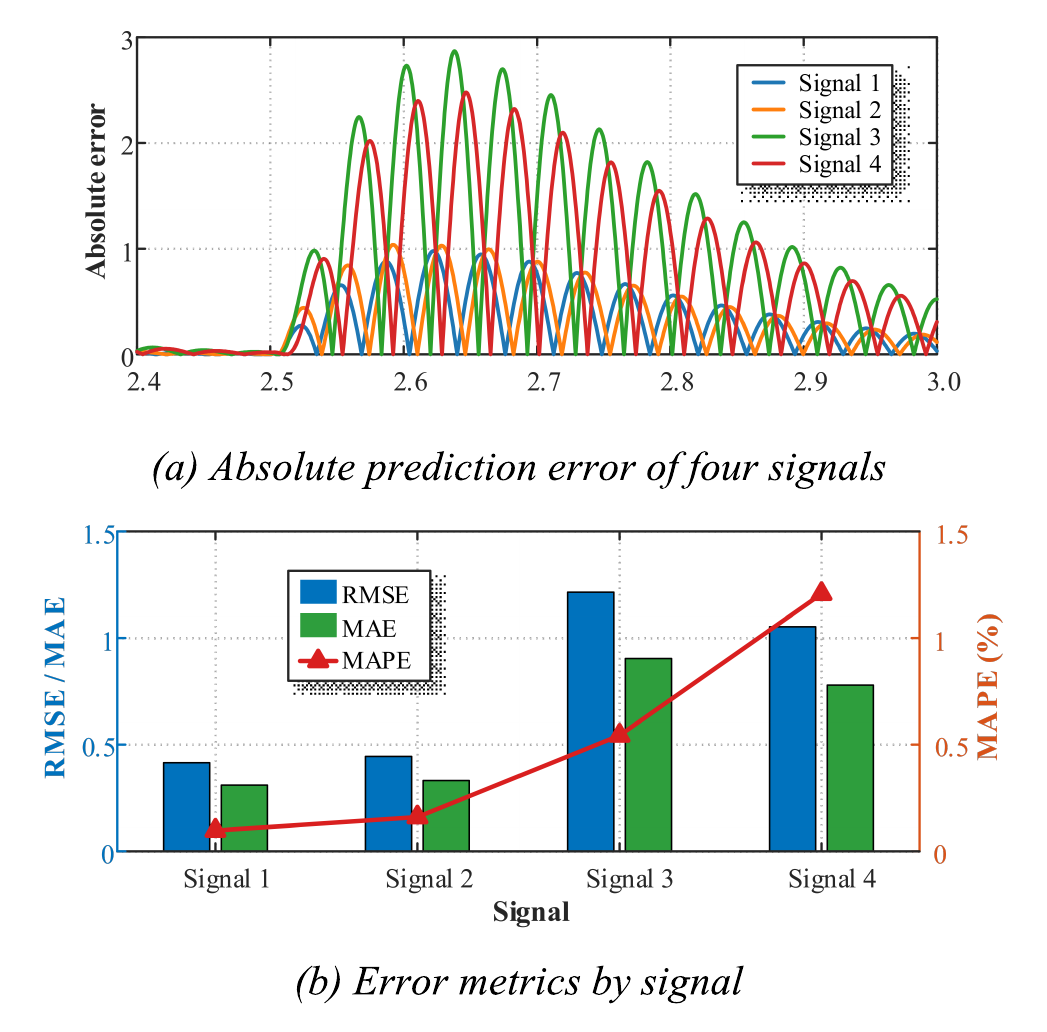}
\caption{Error analysis of system-level substitution simulation in the grid-forming inverter scenario.}
\end{figure}

Fig.~8 further quantifies the substitution error of the grid forming inverter case. The error curves remain small after the disturbance and decay with the recovery of voltage, current, frequency, and power responses. The RMSE, MAE, and MAPE results also show that the main deviations are concentrated around short transient intervals, while the post-disturbance recovery trend is accurately reproduced. This confirms that the proposed method recovers not only offline trajectory fitting relations, but also a closed dynamic model suitable for system-level simulation.

The advantage of MA-LLM-DMD is more evident in this case because the dynamics of a VSG-controlled inverter are jointly determined by LC filter dynamics, power calculation, reference generation, and cascaded voltage and current control laws. Recovering only partial state derivative relations is insufficient for closed-loop simulation. By discovering both differential equations and algebraic closure relations, the proposed method forms an executable model containing state evolution, algebraic variable calculation, and control output generation.

Combining the results in Sections~4.3.3 and 4.3.4, the grid forming inverter case verifies the proposed method from four aspects: reconstruction accuracy, OOD generalization, noise robustness, and system embeddability. Compared with the synchronous generator case, this case further shows that the proposed multi-agent discovery framework can be extended to power electronic dynamic objects with fast electromagnetic transients and complex embedded control logic.

\section{Conclusion}

This study addresses the problem of power system dynamic model discovery under incomplete prior information and proposes a multi-agent collaborative adaptive discovery framework. The framework uses multiple heterogeneous LLM exploratory agents to generate candidate structures in parallel, and combines parameter determination, candidate memory retention, and coordinator broadcasting to realize the joint discovery of differential equations, algebraic closure relations, and key intermediate variables. Case results for synchronous generators and grid-forming inverters show that the proposed MA-LLM-DMD achieves good comprehensive performance in ID/OOD reconstruction accuracy, discovery efficiency, and system-level substitution simulation. Especially in the grid-forming inverter scenario, the proposed method maintains smaller variance and similar ID/OOD errors across multiple runs, indicating that the multi-agent collaboration mechanism can improve structural discovery stability in complex control objects. Future work will further target larger-scale systems and multi-device coupling scenarios, and will study the integration of physical constraints, online updating, and applications for stability analysis and digital twins.


\section*{References}
\begin{thebibliography}{99}
\bibitem{ref1} Shen C, Zuo K, Sun M. Physics-following neural network for online dynamic security assessment[J]. IEEE Transactions on Power Systems, 2025.
\bibitem{ref2} Fabbiani E, Nahata P, De Nicolao G, et al. Identification of AC distribution networks with recursive least squares and optimal design of experiment[J]. IEEE Transactions on Control Systems Technology, 2021, 30(4): 1750--1757.
\bibitem{ref3} Podlaski M, Bombois X, Vanfretti L. Validation of power plant models using field data with application to the Mostar hydroelectric plant[J]. International Journal of Electrical Power \& Energy Systems, 2022, 142: 108364.
\bibitem{ref4} Power E. Handbook of Electrical Power System Dynamics[J].
\bibitem{ref5} Xun Q, Wang P, Li Z, et al. Parameter identification of permanent magnet servo systems based on recursive least squares[J]. Transactions of China Electrotechnical Society, 2016, 31(17): 161--169.
\bibitem{ref6} Virginillo D, Derviškadić A, Paolone M. Identification of power system dynamic model parameters using the Fisher Information Matrix[J]. IEEE Transactions on Power Systems, 2025.
\bibitem{ref7} Asprou M, Kyriakides E. Identification and estimation of erroneous transmission line parameters using PMU measurements[J]. IEEE Transactions on Power Delivery, 2017, 32(6): 2510--2519.
\bibitem{ref8} González-Cagigal M A, Rosendo-Macías J A, Gómez-Expósito A. Parameter estimation of wind turbines with PMSM using cubature Kalman filters[J]. IEEE Transactions on Power Systems, 2019, 35(3): 1796--1804.
\bibitem{ref9} Pereira R F R, De Albuquerque F P, Liboni L H B, et al. Impedance parameters estimation of transmission lines by an extended Kalman filter-based algorithm[J]. IEEE Transactions on Instrumentation and Measurement, 2022, 71: 1--10.
\bibitem{ref10} Zhang Y, Huang X, Xu J. Equivalent mechanical parameter identification strategy for permanent magnet synchronous linear motors based on hybrid adaptive extended Kalman filtering[J]. Proceedings of the CSEE, 2024, 44(03): 1162--1173. DOI:10.13334/j.0258-8013.pcsee.222619.
\bibitem{ref11} Liao G, Wang X. Uncertain model for Thevenin equivalent parameter identification of power systems[J]. Proceedings of the CSEE, 2008, (28): 74--79. DOI:10.13334/j.0258-8013.pcsee.2008.28.012.
\bibitem{ref12} Kontis E O, Skondrianos I S, Papadopoulos T A, et al. Generic dynamic load models using artificial neural networks[C]//2017 52nd International Universities Power Engineering Conference (UPEC). IEEE, 2017: 1--6.
\bibitem{ref13} Pal B C, Coonick A H, Macdonald D C. Robust damping controller design in power systems with superconducting magnetic energy storage devices[J]. IEEE Transactions on Power Systems, 2002, 15(1): 320--325.
\bibitem{ref14} Yang X, Fu Q, Tang B, et al. Grid line parameter identification method based on dynamic spatio-temporal adaptive graph neural networks[J]. Proceedings of the CSEE, 2026, 46(01): 142--156. DOI:10.13334/j.0258-8013.pcsee.241459.
\bibitem{ref15} Brunton S L, Proctor J L, Kutz J N. Discovering governing equations from data by sparse identification of nonlinear dynamical systems[J]. Proceedings of the National Academy of Sciences, 2016, 113(15): 3932--3937.
\bibitem{ref16} Hadifar N, Biglo A H A, Cronin J, et al. Data-Driven Modeling of Power Electronic Converters Using Symbolic Regression for Digital Twin Applications[C]//2025 IEEE Electric Ship Technologies Symposium (ESTS). IEEE, 2025: 135--142.
\bibitem{ref17} Sarić A T, Sarić A A, Transtrum M K, et al. Symbolic regression for data-driven dynamic model refinement in power systems[J]. IEEE Transactions on Power Systems, 2020, 36(3): 2390--2402.
\bibitem{ref18} Shen C, Guo Z, Zuo K, et al. LLM-DMD: Large Language Model-based Power System Dynamic Model Discovery[J]. arXiv preprint arXiv:2601.05632, 2026.
\bibitem{ref19} Shojaee P, Meidani K, Gupta S, et al. LLM-SR: Scientific Equation Discovery via Programming with Large Language Models[J]. arXiv preprint arXiv:2404.18400, 2025.
\bibitem{ref20} Milano F. An open source power system analysis toolbox[J]. IEEE Transactions on Power Systems, 2005, 20(3): 1199--1206.
\bibitem{ref21} Milano F. Power System Analysis Toolbox (PSAT): Documentation for version 2.0.0[R]. University of Castilla-La Mancha, 2006.
\end{thebibliography}
\end{document}